%% file: MainText.tex
\documentclass[12pt,a4paper,oneside]{article}

\usepackage{fourier}
\usepackage[T1]{fontenc}

\usepackage{amsfonts}
\usepackage{amssymb}
\usepackage{amsthm}
\usepackage{amsmath,graphicx,caption,mathtools}
\usepackage{relsize}
\usepackage{fullpage}
\usepackage{subcaption}
\usepackage{booktabs}
\usepackage{changepage}
\usepackage{natbib}
\usepackage[final, bookmarks, bookmarksnumbered=true, bookmarksopen=true, colorlinks=true, hyperfootnotes, linkcolor=black, anchorcolor=blue, citecolor=blue, filecolor=blue, menucolor=blue, runcolor=blue, urlcolor=black]{hyperref}
\usepackage{cleveref}
\usepackage{threeparttable}
\usepackage{tikz}
\usetikzlibrary{automata,positioning,fit, shapes.geometric,calc}
\usepackage{algorithm}
\usepackage[]{algpseudocode}
\usepackage{bbm}
\usepackage{array}
\usepackage{pdfpages}
\usepackage{setspace}
\usepackage{tabularx}
\usepackage{comment}
\usepackage{pdflscape}

\usepackage{dcolumn}
\newcolumntype{d}[1]{D{.}{.}{#1}}
\usepackage{pgfplots}
\pgfplotsset{compat = newest}
\usetikzlibrary{positioning, arrows.meta}
\usepgfplotslibrary{fillbetween}

\newcommand\blfootnote[1]{
  \begingroup
  \renewcommand\thefootnote{}\footnote{#1}
  \addtocounter{footnote}{-1}
  \endgroup
}

\newcolumntype{Y}{>{\centering\arraybackslash}X}

\newtheorem{assumption}{Assumption}

\definecolor{bubbles}{rgb}{0.91, 1.0, 1.0}

\begin{document}

\title{The quality of school track assignment decisions by
teachers}

\author{
Joppe de Ree\thanks{Independent researcher. joppederee@gmail.com}  \and 
Matthijs Oosterveen\thanks{Department of Economics, Lisbon School of Economics and Management, and Advance/CSG, University of Lisbon. oosterveen@iseg.ulisboa.pt} \and
Dinand Webbink\thanks{Department of Economics, Erasmus School of Economics, Erasmus University Rotterdam, Tinbergen Institute, IZA Bonn. webbink@ese.eur.nl
} 
\blfootnote{
We thank Guido Imbens, Hessel Oosterbeek, Nienke Ruijs, Karzan Schippers, Simon ter Meulen, Bas ter Weel, and Thomas van Huizen for valuable comments. 
The paper has also benefited from the comments of participants at the CEP education conference, the LESE conference, the IWAEE conference, the Lisbon School of Economics seminar, and the Utrecht University School of Economics seminar. 
%
%
Dinand Webbink greatly acknowledges receiving a grant from the Netherlands Initiative for Education Research (NRO: project number 405-17-305). The authors have no relevant or material financial interests that relate to the research described in this paper. All omissions and errors are our own. 
Author names are ordered alphabetically.
}}
\date{\today \\ \vspace{0.2in}}

\maketitle
\thispagestyle{empty}

\vspace{-.3in}
\begin{abstract} 
\noindent {\normalsize 
This paper analyzes the effects of educational tracking and the quality of track assignment decisions. We motivate our analysis using a model of optimal track assignment under uncertainty. This model generates predictions about the average effects of tracking at the margin of the assignment process. In addition, we recognize that the average effects do not measure noise in the assignment process, as they may reflect a mix of both positive and negative tracking effects. To test these ideas, we develop a flexible causal approach that separates, organizes, and partially identifies tracking effects of any sign or form. We apply this approach in the context of a regression discontinuity design in the Netherlands, where teachers issue track recommendations that may be revised based on test score cutoffs, and where in some cases parents can overrule this recommendation. Our results indicate substantial tracking effects: between 40\% and 100\% of reassigned students are positively or negatively affected by enrolling in a higher track. Most tracking effects are positive, however, with students benefiting from being placed in a higher, more demanding track. While based on the current analysis we cannot reject the hypothesis that teacher assignments are unbiased, this result seems only consistent with a significant degree of noise. We discuss that parental decisions, whether to follow or deviate from teacher recommendations, may help reducing this noise. 
} 

\medskip
\noindent \textbf{Keywords:} Ability tracking, Student allocation, Regression discontinuity design \\
\textbf{JEL:} D81, I24, C26

\end{abstract}

\setstretch{1} 

\normalsize
\newpage

\setcounter{page}{1}

\input{Introduction.tex}

\input{Setting.tex}

\input{Data.tex}

\input{Thresholdeffects.tex}

\input{Framework.tex}

\input{Results.tex}

\input{Conclusion.tex}

\newpage

\addcontentsline{toc}{section}{References}
\bibliographystyle{chicagoa}
\bibliography{Library.bib}

\linespread{1.00}
\normalsize



\newpage

\appendix

\input{Appendix.tex}

\input{Appendix_results.tex}

\end{document}

%% file: Introduction.tex
\section{Introduction}

Many countries introduce ability tracking in education in some form, and at some point, in the educational careers of students \citep{Betts_2011}. 
There are clear theoretical benefits of tracking. It can increase efficiency by allowing teachers to tailor instruction more precisely to students’ needs, potentially benefiting both high and low achieving students. However, if assignment errors are important and if there are limited opportunities to resolve these errors, or if there are important peer effects, potential benefits from tracking might not materialize on average, or for subpopulations. Prior literature has documented both positive, or at least non-negative, average effects of the supply of tracks \citep{Duflo_etal2011, CardGiuliano_2016, KwakLee_2023}, and negative, or more mixed, average effects of increasing track supply \citep{Piopiunik_2021,Matthewes_2021}.




In this paper we study the quality of track assignment for students at the margin of being assigned to different tracks. The concept of assignment quality has received little attention in the context of educational tracking, whereas biases in the assignment process -- assignment to a track that is not maximizing outcomes in expectation -- could help explain some of the mixed results found in the literature. In particular, using a model of optimal track assignment under uncertainty, we predict that optimal (outcome maximizing) assignment implies, in some of our settings, weakly negative of zero average tracking effects for marginally assigned students. 


The context of our study is the Netherlands, where track assignments are based on a decision process in which teachers first, and parents second, determine the starting track level at which students start secondary education around age 12. There are 5 different secondary school tracks and a variety of mixed, overlapping school tracks that essentially delay the tracking decision by one or two years. 

For the cohorts that we study in this paper, the main factor in the determination of first-year track enrollment in secondary education is the primary teacher's track recommendation. For primary students in 6th grade, teachers determine an initial track recommendation in March of the school year. This initial recommendation is recorded in the administrative systems. In April-May of the school year, students take a standardized school-leavers test.\footnote{There are different suppliers of these tests in the market. These tests should be comparable, but there is discussion about it. In this paper we only look at schools who use the \emph{Cito} school-leavers test. Certainly at the time, the \emph{Cito} test was used by a large majority of primary schools in the Netherlands.} When students score above certain test score cutoffs on this test -- high in the conditional distributions of test scores -- the track recommendation should be formally reviewed by the school. At the review the teacher has to consider an upward revision of the track recommendation and motivate if the upward revision was not applied. In this process, downward revisions are not allowed. 

The revision process of track recommendations allows for a regression discontinuity design (RDD). At specific test-score cutoffs, some -- but not all -- students are reassigned to a higher track. An interesting category of students are those for which the teacher is willing to revise the initial recommendation. For this we consider two types of cutoffs in our data, for which our model predicts different results. At some of these cutoffs the law implies a mandatory reassessment of the initial recommendation. Other thresholds however, which are further up in the test score distribution, mainly act as a ``nudge''. In these settings, crossing a threshold does fundamentally change the regime in which teachers make decisions. Review and potential reassignment can in principle, and does in fact, also occur to students who score just below the cutoff. 
Because reassignment is not mandatory and the teacher’s judgment remains leading throughout the process, we find that in the order of 5\% to 10\% of students are actually reassigned by virtue of a test score just above the cutoff. We find similar effects on teacher reassignment across the various thresholds. 

To analyze the corresponding tracking effects at the cutoffs, we develop a flexible causal approach that is embedded within the context of the RDD. The approach allows for the separation, organization, and partial identification of the various different tracking effects underlying the overall estimated effects at the thresholds. At the test score thresholds, students may experience different types of ``treatments'' effects as there are shifts across more than just two tracks. Moreover, we are interested in separating positive, negative, and total treatment effects of these separate tracks. Indeed, total tracking effects are the sum total of positive and negative effects, which implies that the teacher track assignment process may be ex-post noisy despite being ex-ante unbiased.   


Our results consistently show that tracking has effects for marginal students, with at least 40\% benefiting from reassignment. These effects often persist in the long term. Our theoretical model predicts positive effects at one set of thresholds, while, due to the nature of the decision-making process, it anticipates smaller or even negative effects at some of the other thresholds. The positive effects observed at these thresholds seem largely inconsistent with our model, suggesting that teachers may be assigning conservatively. 


However, since tracking decisions are not made by teachers alone, the behavior of parents and secondary schools can help explain some of these apparent inconsistencies with our model. Parents, for example, may selectively prevent students from accepting upgraded recommendations. This might happen if they expect that starting in a higher track will be harmful. Because we find clear empirical evidence for the phenomenon that starting in a higher track can occur without an upgraded teacher recommendation, we cannot tie the positive tracking effects directly to the upgrading behavior of teachers. This result also prevents us from drawing firm conclusions about whether teacher track recommendations are ex-ante unbiased.

But while we cannot claim much at this point about whether assignment is biased or unbiased, it certainly is noisy. As the starting track matters for many marginally assigned students, one of our policy recommendations is to aim at improving the noisy nature of the current track assignment process in the Netherlands. One option in this context is to potentially involve parents more actively and explicitly in the process of track assignment. Another commonly suggested policy to reduce misallocation is to decrease the number of available tracks altogether. However, our paper shows that this approach risks ``throwing the baby out with the bathwater.'' While it might simplify assignment decisions, it also harms students who, we show in this paper, clearly benefit from from being tracked directly.



One further contribution of our paper is the development of a flexible causal approach, which starts with an extended IV framework. In a setting with one instrument, our IV framework treats the teacher assignment and first-year track enrollment as two multiple ordered treatments in which we aim to identify their separate effects on all margins of a similarly multiple ordered outcome variable. Previous frameworks with one instrument are limited in that they require simplifying treatments into either a single ordered scale \citep{angrist1995two} or a binary variable \citep{imbens1994identification}. Ordering tracks identifies a weighted average of causal effects of unit changes in track enrollment, referred to as the average causal response (ACR). This parameter is difficult to interpret and cannot test our theoretical predictions. Binarizing track enrollment may retrieve the effect of a single track, but the results of our flexible approach demonstrates that in our setting it generates well-known problems with the exclusion restriction \citep{andresen2021instrument}. The inclusion of two treatments -- both teacher assignment and first-year track enrollment -- into our framework is required to test the quality of teacher assignment. Although the specific modeling of these two  treatments is context specific, it does show how an IV framework can accommodate multiple treatments to extract additional insights.\footnote{\cite{nibbering2024instrument,ferman2025dynamic} make a similar point to identify dynamic treatment effects in an IV setting.} Although previous papers often estimate separate treatment effects on the various categories of a discrete outcome (and treatment) variable, they do not wish or need to describe the complete pattern of treatment effects.\footnote{For instance, \cite{Angrist2021Marginal} use randomly assigned financial aid awards to high school graduates to estimate the effect of financial awards on various initial enrollment and degree completion dummies. They subsequently specify and estimate a more parsimonious IV framework that describe degree effects as a function of first-year credits earned only. }

Our causal approach further includes the use of linear programming to partially identify the various causal effects.  In spirit of \cite{imbens1997estimating,abadie2002bootstrap}, we interact the treatment and outcome variable to learn more about the distribution of treatment effects. In particular, we create dummy variables for each value of the treatment and outcome variable, and then construct all possible interactions between these dummies. We estimate the control and treatment means for all these interactions terms. Under the IV assumptions, we link these treatment and control means to the unobserved tracking effects, and there is not more that can be learned about these effects from the data. We use linear programming to retrieve the smallest and largest effect consistent with the estimated control and treatment mean. These bounds are sharp by construction: They are the largest lower and smallest upper bound given the assumptions and what can be identified from the data. 
This approach also relates to the specification test developed by \cite{kitagawa2015test} which also uses interactions between binarized treatment and outcome variables to test necessary conditions of IV validity obtained by \cite{balke1997bounds,imbens1997estimating,heckman2005structural}. Our linear programming approach introduces slack, and if the IV assumptions hold, the program finds a solution while this slack is equal to zero. In contrast, if the IV assumptions are violated, the program will have to make this slack positive to find a solution. We use the degree of slack as a heuristic test for our IV assumptions.

Although previous literature has also considered partial identification approaches in non-standard frameworks, most contributions impose additional assumptions beyond the standard IV assumptions. For instance, \cite{Manski1997MTR} introduces the monotone treatment response (MTR) assumption, which restricts treatment effects to be nonnegative. See, for example, \cite{dehaan2011effect,flores2013partial} for informative applications of the MTR assumption. In our context, however, MTR rules out negative tracking effects, something we are ex-ante unwilling to do considering that teachers assign students to (outcome maximizing) tracks under uncertainty.

%% file: Setting.tex
\section{Setting}\label{sec:Setting}

At the end of primary school, students are assigned to secondary school tracks by primary school teachers in a careful process. The process involves an assessment of performance across multiple primary school years. The primary teacher's track recommendation, as it is referred to, is binding in the sense that students are not permitted to enroll in secondary school tracks above the recommended level. However, perhaps for historical reasons, this rule is not always strictly enforced.  

The Dutch secondary school system broadly consists of five tracks, ranging from the preparatory vocational programs — \emph{vmbo-basis}, \emph{vmbo-kader}, and \emph{vmbo-theoretisch} — to upper general secondary education (\emph{havo}) and pre-university (\emph{vwo}) tracks. Each track grants access to different forms of tertiary education. For example, access to university programs such as law or medicine typically requires a \emph{vwo} diploma.

While a \emph{vwo} diploma is the standard route to university, alternative pathways exist for some programs. For example, students without a \emph{vwo} diploma may first complete a \emph{havo} diploma, then pursue a bachelor's degree at a university of applied sciences (\emph{HBO} – \emph{Hoger beroepsonderwijs}), and subsequently gain access to a university Master's program. However, the feasibility of such routes varies by discipline. For example, medical school almost always requires a \emph{vwo} diploma for access.

\begin{figure}[!hbt]
	\centering
	\caption{\label{fig:trackenrollment4} Secondary school track enrollment by recommended track level, four years after starting secondary education.}
    \includegraphics[width=16cm]{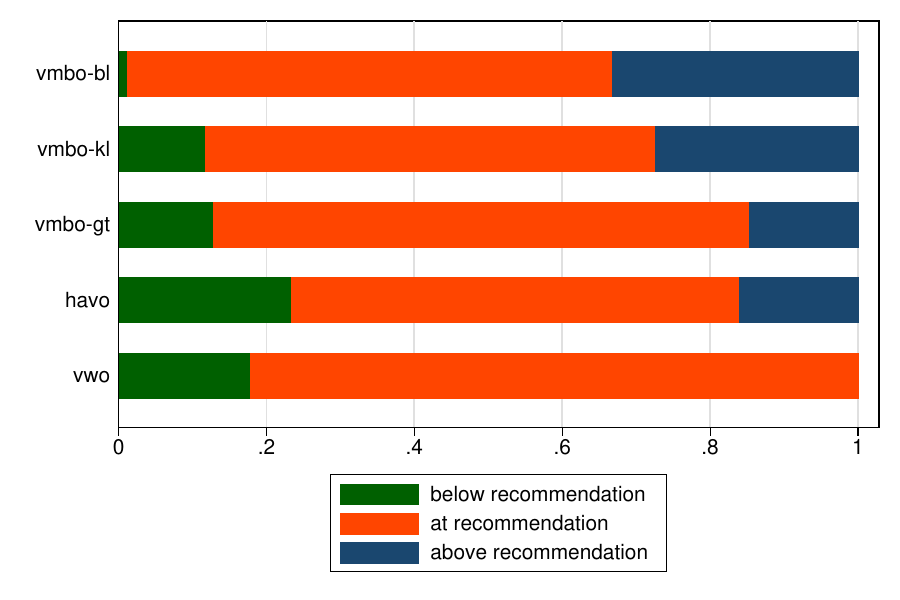}
 \caption*{Notes: The figure is based on almost all students that took the end-of-primary \emph{Cito} test in the 6th grade of primary school in the school years 2014/15, 2015/16 and 2016/17.} 
\end{figure}

The Dutch system also allows for what is known as ``stacking'' of diploma's (Dutch: \emph{stapelen}), whereby students build their educational qualifications sequentially. For example, a student might first earn a \emph{havo} diploma and then transfer into a \emph{vwo} program to obtain a \emph{vwo} diploma that is required for university admission. Figure \ref{fig:trackenrollment4} shows this flexibility of the system in action, by showing the fraction of student below, at or above the track level that was recommended by the teacher. The figure suggests flexibility within the tracked secondary education system. In theory, it is possible to obtain a university diploma by stacking secondary school diploma's. On the other hand, stacking requires a level of commitment and perseverance from students that may not be given to everyone. 

In this paper we study the effects of tracking by evaluating the medium and longer term effects of enrolling in different tracks. For this we use a feature of the Dutch track assignment process that was introduced in the school year 2014/15. In 2014/15, the regulations of the track assignment process were changed in two ways. First, the primary school teacher's recommendation became binding for secondary education track placement \citep{wvo2014}. Secondary schools were not (as a rule) permitted to place students on track levels above the recommended level. Second, an option was introduced for the primary teachers to upwardly revise the track recommendation, based on a test scores above specific test score cutoffs on the standardized end-of-primary education test.

For the period that we are studying, the procedure of track recommendations proceeded as follows. In March of the the school year, all students would receive a secondary school track recommendation. The recommended level might be one of six track levels, from practice-based secondary education (\emph{praktijkonderwijs}) to the pre-university \emph{vwo}. Mixed, or combined recommendations, such as \emph{havo}/\emph{vwo} are also possible, and are, in fact, quite common. These track recommendations are formally recorded in the administrative systems of the \emph{Dienst Uitvoering Onderwijs}. 

In April or May of that school year, students take the standardized end-of-primary education test. One purpose of this test is to provide a ``second opinion'' to the teacher's recommendation. These achievement test scores map into \emph{suggested} track levels, based on nationwide and predetermined test score cutoffs. We refer to the mapping from the test score to these suggested track levels as the \emph{test-based recommendation}. If a student's \emph{test-based recommendation} exceeds the track level that was initially recommended in March, the teacher must formally reassess the initial recommendation. Throughout the process, however, the teacher's motivated opinion remains leading. The reassessment therefore does not automatically translate into a revised track recommendation. Teachers may, as permitted by law, refrain from an upgrade. In such cases, however, they are required to provide a motivation for this decision.\footnote{For context, we refer to article 42 of the law on primary education (\url{https://wetten.overheid.nl/BWBR0003420/2016-01-18/0})}

\begin{table}[!htb]
	\centering
	\caption{\label{tab:structure_upgrading} Schematic presentation of the track recommendation upgrading procedure}
    \includegraphics[width=\textwidth]{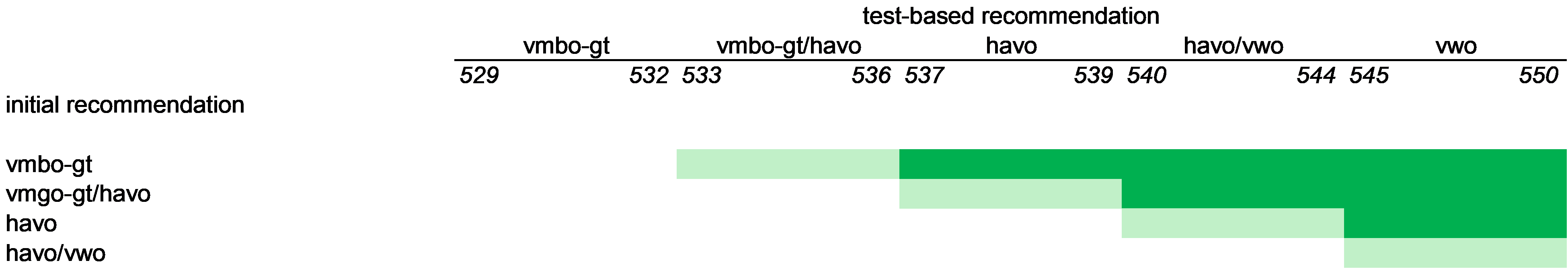}
 \caption*{\emph{Notes}: In Appendix \ref{app:Cutoffs} we present the complete mapping from test scores to test-based recommendation, including all the cutoff levels, for all the cohorts in the data.}
\end{table}

In Table \ref{tab:structure_upgrading} we show schematically how the upgrading process works. In rows we present a selection of the initial recommendations that are possible. The top rows indicate the mapping from a test score (in brackets) to a \emph{test-based recommendation}. For example, a test score of 530, falls into the first bracket in the table, and implies a \emph{test-based recommendation} of \emph{vmbo-gt}. For the initial recommendations indicated in the first column, a test score of 530, or in other words, a test-based recommendation of \emph{vmbo-gt}, has no formal implications. When the \emph{test-based recommendation} is at, or below the level of the initial recommendation, primary teachers do not need to review their initial recommendations. Having said that, upgrading students, by recommending a higher track level than the initial recommendation, is always possible in response to any test result.  

The green areas indicate test based-recommendations, that would translate into a reassessment of the initial recommendation. For example, for students with an initial \emph{vmbo-gt} recommendation, the recommendation must be reassessed with a test-based recommendation of \emph{vmbo-gt/havo} or higher. For students with an initial \emph{vmbo-gt} recommendation, this occurs with a test score of 533 or higher. For students with higher level initial recommendations, this first relevant threshold, is shifted to the right. For students with an initial \emph{havo} recommendation, the first relevant threshold, for which a reassessment is mandated, is at a score of 540. 

It is important to mention that, conditional on the decision to upgrade the recommendation, teachers are not obligated to assign the exact level indicated by the test-based recommendation. That is, teachers may upgrade to any other level they prefer, provided that it is above the level of the initial recommendation. In some cases, the test-based recommendation provides only a nudge. However, it is potentially not a pure nudge in the behavioral economics sense, as deviating from it (by not upgrading or by upgrading to a lower level than the level indicated by the test-based recommendation) also requires justifying that choice to parents, who might oppose it. This adds implicit costs, making the test-based recommendation more than a neutral signal.

In Section \ref{sec:model} we work out a simple descriptive model in an attempt to be concrete about the different elements that might affect incentives of whether and how to upgrade the initial track recommendation. We then use this model to make predictions about who is reassigned and what kind of longer term effects we can expect from these reassignments. In the model, we separately consider the white regime vs. the light green regime and the light green regime vs. the dark green regime. The predictions of this model play a key role in the interpretation of our findings in \Cref{sec:results}.

\subsection{Model}\label{sec:model}

Student assignment is influenced by different institutional incentives across white, light green, and dark green regimes. In the white regimes, teachers may face soft pressures to assign students conservatively. As secondary schools face scrutiny from the education inspection for excess downward track mobility, overly ambitious recommendations might create a problem for them. In the light green and dark green regimes, these incentives disappear because the inspection does not consider the upgraded recommendations when they calculate track mobility. 

Another key distinction between the white and light green regimes is, of course, that in the latter, teachers are required to reassess their initial recommendation. This potentially leads to more accurate assignments. Between the light green and the dark green regime there is no such asymmetry, reassessment takes place on both sides of the cutoffs. Instead, between the light and dark green regimes the nudge of the track based recommendation might play a role, just as non-monitory (psychological/emotional) costs of having to justify to parents and students that they do want to upgrade the initial recommendation. 

In the remainder of this section we work out a theoretical model for a two-track case, to further develop intuition. The main component of this model is a preference for (binary) high track attainment after four years of secondary education $H_4$. In the various regimes, specific incentives might play a role that induce teachers to deviate from just maximizing the likelihood that students reach $H_4=1$.

Suppose teachers maximize the following expected utility function, by choosing $H_0=1$ (a high track recommendation) or $H_0=0$ (a lower track recommendation):
\begin{align*}
&E[U(H_0)|I_B+I_T\times (1-Z_{white})]  = \\
&\beta E[H_4(H_0)|I_B+I_T\times (1-Z_{white})] - \gamma Z_{white} H_0 - \delta 1(H_0<TBR(Z))
\end{align*}
where $H_4$ is a function of $H_0$. The expectation is conditioned on $I_B+I_T\times (1-Z_{white})$, where $I_B$ is the information used for the initial recommendation and $I_T$ is the information available to decide on the upgrade. $Z_{white}=1$ indicates the white regime. Hence, it is assumed in this model that the information $I_T$ is not considered in the white regime (although it is available to them). The term $TBR(Z)$ indicates the level of the test-based recommendation.

The utility specification consists of three components:
\begin{itemize}
\item[C1.] The $\beta E[H_4(H_0)|I_B+I_T\times (1-Z_{white})]$ is the utility value of the likelihood of reaching $H_4=1$ as a result of the teacher's decision $H_0=1$ or $H_0=0$. The expectation is conditioned in information that can be incorporated in the decisions, which might differ between the (light and dark) green regimes on the one hand, and the white regime on the other. While teachers are always allowed to upgrade recommendations, even in the white regime, they might choose not to incorporate the new information of the test $I_T$ in their decision-making.
\item[C2.] The $-\gamma Z_{white} H_0$ indicates a negative utility value for assigning high, in the $Z_{white}=1$ regime. This is modeling the idea that there might be soft incentives to assign conservatively. 
\item[C3.] The $- \delta 1(H_0<TBR(Z))$ measures the non-monetary (psychological) cost of  recommending a track level below the $TBR(Z)$ as well as a nudge, provided by the \emph{test-based recommendation}. These \emph{test-based recommendations} naturally depend on the assignment regime $Z$ (where $Z$ may be $Z_{white}$, $Z_{light green}$ and $Z_{dark green}$.
\end{itemize}
Based on this model, we can derive predictions about who will be upgraded at the the various thresholds and, to the extent that teachers form rational expectations, what the effects of these upgrades might be. 

Upgrading at the thresholds between the regimes takes place if on the left side of the threshold $H_0=0$ is selected and on the right side of the threshold $H_0=1$ is selected. In general, we can derive that $H_1=1$ is selected when:
\begin{align*}
    &\beta E[H_4(1)-H_4(0)|I_B+I_T\times (1-Z_{white})] - \gamma Z_{white} + \delta 1(0<TBR(Z))>0
\end{align*}

\subsubsection{The white vs. the light green regime}

In white $H_0=0$ is selected if:
\begin{align}
\beta E[H_4(1)-H_4(0)|I_B] < \gamma 
\end{align}
In light green $H_0=1$ is selected if:
\begin{align}
\beta E[H_4(1)-H_4(0)|I_B+I_T] + \delta > 0 
\end{align}
We further define:
\begin{align}
    E[H_4(1)-H_4(0)|I_B+I_T] = E[H_4(1)-H_4(0)|I_B] + \nu_T
\end{align}
where $\nu_T$ measures the extent to which the expectation $E[H_4(1)-H_4(0)]$ has increased or decreased by virtue of the available information $I_T$. The $\nu_T>0$ if the new information makes teachers lean more towards the high track recommendation $H_0=1$.

Upgrading occurs at a shift between the two regimes if both conditions hold at the same time:
\begin{align}
-\delta < \beta E[H_4(1)-H_4(0)|I_B+I_T]< \gamma +\beta \nu_T 
\end{align}
Among those who are reassigned we might find some slightly negative effects because of the $\delta$, which permits that $E[H_4(1)-H_4(0)|I_B]$ can be smaller than 0. Due to the $\gamma$ parameter, $E[H_4(1)-H_4(0)|I_B]$ is also permitted to be larger than zero. Between the white and light green regimes we might also find positive effects of reassignments. Similarly, if the $\nu_T$ term is positive, positive effects of the upgrade, for those who are reassigned are permitted in this model. 

\subsubsection{The light green vs. the dark green regime}

In light green assignment to $H_0=0$:
\begin{align}
\beta E[H_4(1)-H_4(0)|I_B+I_T] < 0
\end{align}
In dark green assignment to $H_0=1$:
\begin{align}
\beta E[H_4(1)-H_4(0)|I_B+I_T] + \delta > 0
\end{align}
Upgrading occurs when both conditions hold:
\begin{align}
-\delta < \beta E[H_4(1)-H_4(0)|I_B+I_T] < 0
\end{align}
At the light green vs. dark green threshold, rational expectations suggests small and occasionally negative effects. These effects are driven by upgrades resulting from the nudging role of the \emph{test-based recommendation}, as well as the psychological costs teachers face when having to justify to parents and students why they choose not to upgrade an initial (low) recommendation despite a high test score. Indeed, taking such decisions requires a considerable level of confidence in the ability to make such judgments.

Before we continue with the empirical sections, it is important to also consider the role of parents and students, as well as other more random aspects of the assignment process, such as placement restrictions. While teachers might upgrade the recommendations at these thresholds, parents (and students) might not take full advantage of this opportunity. Parents always have the option to enroll at a level below the recommended level. This may be appealing to parents and students if they believe that a lower track level is the expected outcome maximizing track for them. The consequence of this two-step decision-making process is that the observed effects of a change in the assignment regime might not be driven by a change in the track recommendation alone. It is possible that parents have additional information and are able to correct some of the potential errors teachers might make in their assignments. The predictions presented in this section, then, might no longer hold exactly. We return to this in the Section \ref{sec:results} when discussing our findings.

%% file: Data.tex
\section{Data}


We use proprietary administrative data from Statistics Netherlands on all students that take the standardized end-of-primary education test in the 6th grade of primary school in the school years 2014/15 until 2018/19. We refer to these three different groups of students as cohorts.
The 2014/15 cohort is the first that is affected by the new track assignment regulations discussed in Section \ref{sec:Setting}. 
Our main outcome variable tracks students four years into secondary education. Our longer-term outcome variables follow students up to eight years through secondary and tertiary education, for which we only use the first three cohorts until 2016/17.

For almost all students we observe the initial teacher track recommendation, the scores on the standardized end-of-primary education test, and the potentially revised track recommendation. For all five cohorts, we also observe track enrollment in the first four years of secondary education, and their corresponding major choice (Dutch: \emph{Profielkeuze}).\footnote{In the third or fourth year, depending on their track enrollment, students have to choose a major that greatly affects their future coursework in secondary education and their options in higher education.} We follow the first three cohorts, 2014/14 until 2016/17, for eight years through secondary and tertiary education. For secondary education, we record the  highest completed track. If a student has not yet graduated after eight years, we record their latest track enrollment instead. For tertiary education, we register the highest level of enrollment observed within the eight-year follow-up period. We also observe several relevant background characteristics, including gender, age, and household income.

%

There are two criteria that we use to construct our final sample. First, our final sample only contains students from primary schools that use the end-of-primary education test provided by test developer \emph{Cito}. While \emph{Cito} is still by far the largest provider of this test, other test developers have more recently entered the market for these tests. 
%
Second, we select the students who start secondary education in the year after they are assigned. That is, for students who repeat the 6th grade of primary school, we use the last observed enrollment in grade 6.

%% file: Thresholdeffects.tex
\section{Threshold effects}\label{sec:threshold_effects}

In the next section \ref{sec:framework} we develop a flexible causal approach in an attempt to characterize and estimate the quantities we need to make an assessment of the quality of track assignment. Key features of this methodology are that we want to assess the average effects of reassignment, but also, as a marker of noise in the decision-making process, something we call the total effect of a reassignment. The total effect is the sum  of positive and negative effects. 

We also want to disentangle these effects for different shifts in the track enrollments. As we have mentioned before, the Dutch secondary school system consists of many different tracks, which some track types overlapping others. Particularly, with results of \cite{Duflo_etal2011} in mind, an upgrade might have (unforeseen) positive and negative average effects at the same time. For example, in the Netherlands hybrid tracks which provide education at the level of two (or more) different tracks. The purpose of these hybrid tracks is essentially delay the tracking decision by some years. It is possible that some students would shift into such a hybrid track, e.g. from \emph{havo} to \emph{havo/vwo}, while other shift away from it, from \emph{havo/vwo} to \emph{vwo}. 


Prior to developing this approach we first present some of our data in a more straightforward and conventional way. In this section we present simple comparisons left and right of the relevant thresholds, on some of the outcomes of interest. This conventional presentation of the data is not flexible enough to answer some of the more precise policy questions that we are interested in. The simple causal comparisons however show clear first evidence of the existence of average tracking effects, which also persist into the long term.

For the empirical results in this section, as well as later in Section \ref{sec:results}, we partition outcomes (tracks, or other indicators of educational attainment) in three groups: Low ($L)$, Middle ($M$), and High ($H$). In most of the specifications we will look at tracks, whether these tracks are recommendations or actual enrollments. 

The binary outcomes might indicate the (final) track recommendation ($L_0$, $M_0$ and $H_0$), track enrollment in the first year of secondary education ($L_1$, $M_1$ and $H_1$), and track enrollment four years after starting secondary education ($L_4$, $M_4$ and $H_4$).  
The level of the initial recommendation is always indicated by $L_0=1$. Subsequently, $M_0=1$ and $H_0=1$ always indicate a half and whole step (or more) up in the ladder of track recommendations. For example, for students with an initial \emph{havo} recommendation, $L_0=1$ indicates a \emph{havo} recommendation, $M_0=1$ indicates a \emph{havo/vwo} recommendation and $H_0=1$ indicates a \emph{vwo} recommendation. 

For the enrollments and for longer term outcomes, the mapping of specific tracks to the placeholders $L$, $M$ and $H$ are based on relevant categories. In particular, for example, for outcomes, we consider a ``Low'' category that is particularly ``Low'', given the initial recommendation. We do this, because we want to specifically allow for the possibility that upgrading has negative effects on track enrollment after four years of secondary education, or even later in tertiary education. We present the full mapping of all the optional tracks to the outcomes $L_0$, $M_0$, $H_0$, $L_1$, $M_1$, $H_1$, $L_4$, $M_4$, and $H_4$ in Appendix \ref{app:coding_rules}.

\subsection{Regression discontinuity design}


Let $L_t$, $M_t$, and $H_t$ be the corresponding three mutually exclusive and exhaustive track dummies in year $t$. With $t=0$ the dummy variable refers to the teacher recommendation, where for instance $L_0$ is equal to one if the teacher does not upgrade. With $t=1$ and $t=4$ the dummy variable refers to year one and year four track enrollment respectively. For instance, $M_1$ and $H_4$ are equal to one if the student enrolls in the middle track in year one, and the high track in year four, relative to the initial track recommendation. Throughout the paper we suppress the student index for notational convenience.

Let $S$ be the test score centered at the cutoff. We define the threshold effect ($\tau_Y$) as the difference in the average outcomes between students just above and just below the test score cutoff, 
\begin{align*}
   \tau_Y =& \underset{s \to^+0}{\lim}\mathbb{E}[ Y|S=s]-\underset{s \to^- 0}{\lim}\mathbb{E}[Y|S=s].
\end{align*}
The variable $Y$ is one of the nine outcome variables, namely one of the three track dummies across the three years. This mimics a standard regression discontinuity design (RDD), and our threshold effects are ``reduced-form'' effects of scoring just above the cutoff. Heuristically, these effects identify the causal effect of scoring just above the cutoff if students left and right of the cutoff are similar ex-ante.

Our empirical implementation to the estimation of threshold effects follows the literature. In particular, we estimate the following RD model:
\begin{align}
\label{eqn:rf}
    Y=\alpha_{Y}+\beta_{Y} Z + f_{Y}(S) + \epsilon_{Y}, 
\end{align}
where the dummy variable $Z$ is equal to one if the student scores above the test score cutoff:
\begin{align*}
    Z = \left\{
    \begin{array}{ll}
        0 & S < 0, \\
        1 & S \geq 0.
    \end{array}
\right.
\end{align*}

The polynomial $f(S)$ and bandwidth are important considerations for the RD model in \eqref{eqn:rf}. Following \cite{imbens2008regression,cattaneo2020foundations}, we specify a polynomial of degree one that is allowed to differ on each side of the cutoff. The \emph{Cito} test score ranges from 501 to 550, and therefore has a discrete set of 50 points.  We use a symmetric bandwidth of three test score points on both sides of the cutoff. This bandwidth aligns well with results from the several data-driven bandwidth selection procedures using the Stata command \emph{rdrobust} proposed by \cite{calonico2017rdrobust}. We estimate this RD model using OLS with robust standard errors \citep{kolesar2018inference}, separately for each of the initial track assignments and cutoffs. We test for the robustness of our results using a polynomial of degree two and a symmetric bandwidth of two and four test score points, and also report standard errors based upon 1000 bootstrap samples. 

The threshold effect is estimated by $\beta_Y$. We will also show the average outcome of students just below the test score cutoff, which is estimated by $\alpha_Y$, and referred to as the control mean (cm). 
Note that our fixed bandwidth across outcome variables ensures that our estimates for $\beta_Y$ ($\alpha_Y)$ exactly add up to zero (one) across the three track dummies in one year.

\subsection{Baseline results}\label{sec:baseline_results}

\begin{figure}[!htb]
	\centering
	\caption{\label{fig:RD_60} Fraction of students by achievement score, with a mixed \emph{havo}/\emph{vwo} recommendation [A] and a \emph{vwo} recommendation [B], for the sample of students with an initial \emph{havo} recommendation}
    \includegraphics[width=15cm]{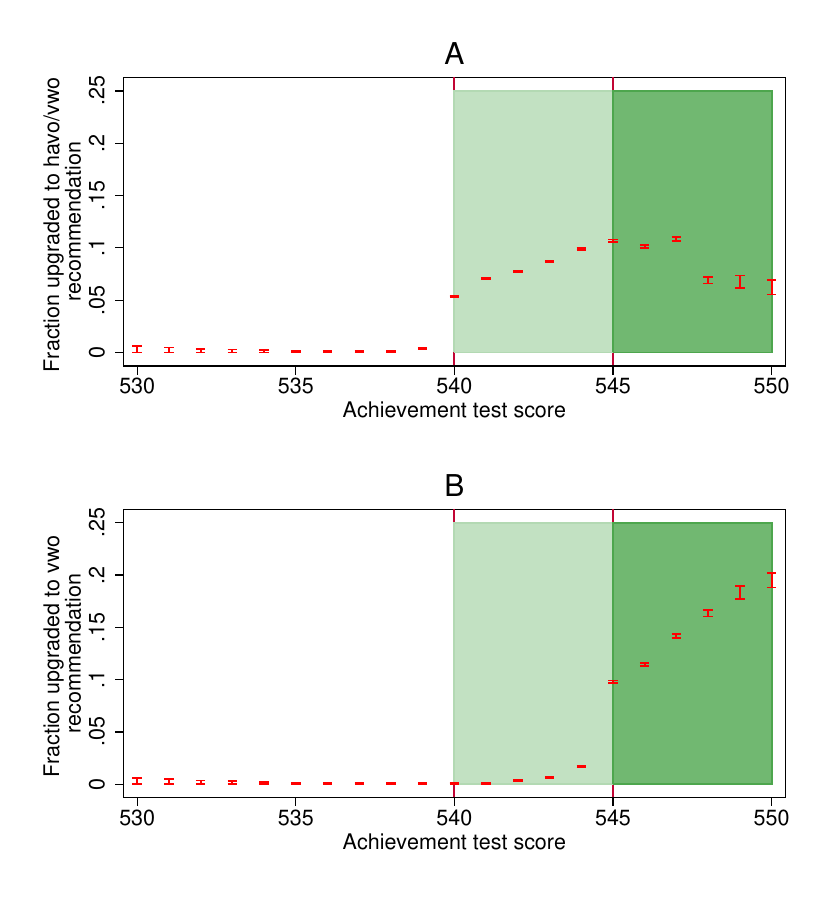}
 \caption*{\emph{Notes}: The Figure shows ranges of fractions instead of point estimates at each test score level. This is due to privacy restrictions for using this data. These limitations do not apply in the same way to the rest of the quantitative results in this paper.}
\end{figure}

In a first empirical step, we focus on the estimate $\beta_Y$, conditional on a particular initial track recommendation. To get a sense of what our results look like graphically, we zoom in on students with an initial \emph{havo} recommendation in Figure \ref{fig:RD_60}. Figure \ref{fig:RD_60}A presents the fraction of students with a $M_0=1$ recommendation, which is here a \emph{havo/vwo} recommendation. The Figure \ref{fig:RD_60}B presents the fraction of students with a $H_0=1$ recommendation, which is a \emph{vwo} recommendation. The figure also clarifies the different assignment regimes, using the same color coding as we used in Table \ref{tab:structure_upgrading}. 

The Figure conveys a lot of preliminary information. For example, at the relevant thresholds, teachers tend to upgrade. But, as it turns out, teachers can only be moderately motivated to upgrade at the relevant thresholds. Crossing the threshold level, from the white to the light green regime yields an effect on receiving a \emph{havo/vwo} recommendation of about 6\%. For effects on receiving a  \emph{vwo} recommendation, we need to look at higher test score values. Crossing the threshold from the light green into the dark green regime, yields an effect on receiving a \emph{vwo} recommendation of about 8\%. 

At the threshold between the light green and the dark green regime, at the 545 score, the figure also suggest that multiple differential treatments might occur at the same time. At this threshold there is no effect on the \emph{havo/vwo} recommendation, while there is a strong effect on the \emph{vwo} recommendation. One possible explanation of this result is that, students are upgraded only from the initial \emph{havo} recommendation (directly) to \emph{vwo}, without any of them receiving the intermediate \emph{havo/vwo} recommendation. Another, in our view more plausible explanation for this is that a (nonzero) number of students are upgraded from \emph{havo} to a \emph{havo/vwo} and an equal number of students from \emph{havo/vwo} to \emph{vwo}. The possibility that different students receive different treatments, at the same threshold, plays an important role in the next sections, were we attempt to disentangle them. 

In Table \ref{tab:thresholdeffects_E0} we present estimates of the threshold effects on $L_0$, $M_0$ and $H_0$, for all of the optional initial recommendations separately. We also consider two thresholds for each of the initial recommendations. We consider the shift from the white to the light green regime (indicated with $+ 1$), and from the light green to the dark green regime (indicated with $+ 2$). The results shown in the rows ``havo $+ 1$'' and ``havo $+ 2$'' are based on the same data used for Figure \ref{fig:RD_60}A and B respectively. 

\begin{table}[!htb]
\begin{singlespace}
    \begin{center}
    \begin{threeparttable}
    \caption{\label{tab:thresholdeffects_E0} Threshold effects on recommended track level}
	\begin{tabularx}{1\textwidth}{@{\extracolsep{\fill}} lcccccc}
		\toprule
         &\multicolumn{2}{c}{Low ($L_0$)}&\multicolumn{2}{c}{Middle ($M_0$)}&\multicolumn{2}{c}{High ($H_0$)} \\ 
        &$\alpha_Y$&$\beta_Y$&$\alpha_Y$&$\beta_Y$&$\alpha_Y$&$\beta_Y$\\    
            \cmidrule(lr){2-7}
            &(1)&(2)&(3)&(4)&(5)&(6)\\
		\cmidrule(lr){2-7}
    \input{Tables2/table_rf_60w_baseline_Y0.csv}\\
    \input{Tables2/table_rf_52w_baseline_Y0.csv}\\
    \input{Tables2/table_rf_50w_baseline_Y0.csv}\\
    \input{Tables2/table_rf_34w_baseline_Y0.csv}\\
    \input{Tables2/table_rf_30w_baseline_Y0.csv}\\
    \input{Tables2/table_rf_22w_baseline_Y0.csv}\\
    \input{Tables2/table_rf_20w_baseline_Y0.csv}\\
    \input{Tables2/table_rf_61h_baseline_Y0.csv}\\
    \input{Tables2/table_rf_60h_baseline_Y0.csv}\\
    \input{Tables2/table_rf_52h_baseline_Y0.csv}\\
    \input{Tables2/table_rf_50h_baseline_Y0.csv}\\
    \input{Tables2/table_rf_34h_baseline_Y0.csv}\\
    \input{Tables2/table_rf_30h_baseline_Y0.csv}\\
    \input{Tables2/table_rf_22h_baseline_Y0.csv}\\
    \input{Tables2/table_rf_20h_baseline_Y0.csv}\\
    \bottomrule
	\end{tabularx}
\begin{tablenotes}[flushleft]
\item \emph{Notes.} ***, **, * refers to statistical significance at the 1, 5, and 10\% level. Robust standard errors for estimates of $\beta_Y$ in parentheses. The table shows estimated parameters $\alpha_Y$ and $\beta_Y$ (as presented in equation \ref{eqn:rf}) on recommended track levels Low ($L_0$), Middle ($M_0$) and High ($H_0$).
\end{tablenotes}
\end{threeparttable}
\end{center}
\end{singlespace}
\end{table}

The Table \ref{tab:thresholdeffects_E0} shows significant upgrading at each of the thresholds in column (2). Only moderately depending on the setting, we find that about 10\% is upgraded at the thresholds, as students just to the right of the thresholds, are less likely to still have the initial track recommendation they had received in March. This is a first key result of the paper. At each of these thresholds there is upgrading. But, as it turns out, teachers are only moderately willing to upgrade the recommendation to a higher track level. This result, in our view, indicates that teachers are generally quite confident in their decisions. It indicates also that teachers tend to take their own professional judgment on the assignment of graduating primary students to secondary school tracks very seriously. 

The relatively low rates of upgrading is in our view also not really surprising. Primary students in the Netherlands are subject to a rigorous testing regime. From first grade onward, they are assessed biannually in math and language using high-quality, nationwide standardized tests. These assessments allow teachers to closely monitor each student’s academic development and compare their performance to national benchmarks. In addition to this, teachers also rely on their own professional judgment and inputs from colleagues (including teachers from earlier grades). Given this comprehensive information on students ability and achievement, any single (high) score on the school-leavers test might not provide much new information. 

For interpreting the effects on the upgrade, we refer to Section \ref{sec:model} in the previous section. At the $+ 1$ thresholds, the upgrade might reflect one of a variety of motivations. At the $+ 2$ thresholds, there are fewer theoretical arguments for upgrading. The fact that we see upgrading across the board suggest that teachers are feeling some pressure to assign the test-based recommendation, and that they are sensitive to the nudge it provides. Both might work hand in hand, when they are willing to provide the upgrade in about 10\% of the cases, when they are not too strongly opposed to it. On the other, they also cannot be strongly supporting it, because in that case they could just as easily provide the same upgrade to the left of the $+ 2$ threshold. 

If the change in the recommendation is mapped fully into a change in enrollment in the first year of secondary education, we might be able to draw quick conclusions about the assignment quality of teachers, based on the model presented in Section \ref{sec:model}. In Table \ref{tab:thresholdeffects_E1} however we show that enrollment effects are not the same as the effects on the recommendation we have seen in Table \ref{tab:thresholdeffects_E0}. In essence, we find that the enrollment effects are weaker than the effects on the recommendation.\footnote{As the choice set for enrollment is different than the relevant ranges of the recommended level, the definitions for $L_1$, $M_1$ and $H_1$ do not align perfectly with the definitions for $L_0$, $M_0$ and $H_0$. See Appendix \ref{app:coding_rules} for the exact mappings from tracks to these placeholders $L$, $M$ and $H$.} This is due to the fact parents and students have an independent decision to make. Parents might for example disagree with teachers about which track is the expected outcome maximizing track.



Based on the arguments presented in Section \ref{sec:model} we anticipate different effects on outcomes at the $+1$ and $+2$ thresholds. At the same time, Table \ref{tab:thresholdeffects_E1} shows that we cannot ignore the role of parents and students in the decision-making process. Model predictions discussed in Section \ref{sec:model} now seems to require an additional layer of complexity, which we will aim to accommodate in the causal framework of \Cref{sec:framework}.

\begin{table}[!htb]
\begin{singlespace}
    \begin{center}
    \begin{threeparttable}
    \caption{\label{tab:thresholdeffects_E1} Threshold effects track enrollment in first grade secondary}
	\begin{tabularx}{1\textwidth}{@{\extracolsep{\fill}} lcccccc}
		\toprule
         &\multicolumn{2}{c}{Low ($L_1$)}&\multicolumn{2}{c}{Middle ($M_1$)}&\multicolumn{2}{c}{High ($H_1$)} \\
    &$\alpha_Y$&$\beta_Y$&$\alpha_Y$&$\beta_Y$&$\alpha_Y$&$\beta_Y$\\    
            \cmidrule(lr){2-7}
            &(1)&(2)&(3)&(4)&(5)&(6)\\
		\cmidrule(lr){2-7}
    
    \input{Tables2/table_rf_60w_baseline_Y1.csv}\\
    \input{Tables2/table_rf_52w_baseline_Y1.csv}\\
    \input{Tables2/table_rf_50w_baseline_Y1.csv}\\
    \input{Tables2/table_rf_34w_baseline_Y1.csv}\\
    \input{Tables2/table_rf_30w_baseline_Y1.csv}\\
    \input{Tables2/table_rf_22w_baseline_Y1.csv}\\
    \input{Tables2/table_rf_20w_baseline_Y1.csv}\\
    \input{Tables2/table_rf_61h_baseline_Y1.csv}\\
    \input{Tables2/table_rf_60h_baseline_Y1.csv}\\
    \input{Tables2/table_rf_52h_baseline_Y1.csv}\\
    \input{Tables2/table_rf_50h_baseline_Y1.csv}\\
    \input{Tables2/table_rf_34h_baseline_Y1.csv}\\
    \input{Tables2/table_rf_30h_baseline_Y1.csv}\\
    \input{Tables2/table_rf_22h_baseline_Y1.csv}\\
    \input{Tables2/table_rf_20h_baseline_Y1.csv}\\
    \bottomrule
	\end{tabularx}
\begin{tablenotes}[flushleft]
\item \emph{Notes.} ***, **, * refers to statistical significance at the 1, 5, and 10\% level. Robust standard errors for estimates of $\beta_Y$ in parentheses. The table shows estimated parameters $\alpha_Y$ and $\beta_Y$ (as presented in equation \ref{eqn:rf}) on first year secondary school track enrollment levels Low ($L_1$), Middle ($M_1$) and High ($H_1$).
\end{tablenotes}
\end{threeparttable}
\end{center}
\end{singlespace}
\end{table}

The fact that we have been able to document changes in first year track enrollments, also suggests an opportunity to study  the effects that different enrollments might have on outcomes. 
In Table \ref{tab:thresholdeffects_E4} we present results on track enrollment after four years of secondary education. Note that for fourth-year track enrollment the Low category is Lower than for teacher assignment and first-year enrollment.  Generally, we find positive effects on these medium term outcomes. These effects are also considerably large, suggesting that a large share of the students who are upgraded and/or enroll in a higher track in first year, benefit from this in the medium term. For example, for student with an initial \emph{havo} recommendation, who are reassigned at the $+ 2$ threshold, we estimate that 4.2\% is enrolled at \emph{vwo}, who would otherwise be enrolled at a lower level, most likely (but not certainly) \emph{havo}. 

\begin{table}[!htb]
\begin{singlespace}
    \begin{center}
    \begin{threeparttable}
    \caption{\label{tab:thresholdeffects_E4} Threshold effects track enrollment after four years of secondary education}
	\begin{tabularx}{1\textwidth}{@{\extracolsep{\fill}} lcccccc}
		\toprule
         &\multicolumn{2}{c}{Low ($L_4$)}&\multicolumn{2}{c}{Middle ($M_4$)}&\multicolumn{2}{c}{High ($H_4$)} \\
&$\alpha_Y$&$\beta_Y$&$\alpha_Y$&$\beta_Y$&$\alpha_Y$&$\beta_Y$\\    
            \cmidrule(lr){2-7}
            &(1)&(2)&(3)&(4)&(5)&(6)\\
		\cmidrule(lr){2-7}
    
    \input{Tables2/table_rf_60w_baseline_Y4.csv}\\
    \input{Tables2/table_rf_52w_baseline_Y4.csv}\\
    \input{Tables2/table_rf_50w_baseline_Y4.csv}\\
    \input{Tables2/table_rf_34w_baseline_Y4.csv}\\
    \input{Tables2/table_rf_30w_baseline_Y4.csv}\\
    \input{Tables2/table_rf_22w_baseline_Y4.csv}\\
    \input{Tables2/table_rf_20w_baseline_Y4.csv}\\
    \input{Tables2/table_rf_61h_baseline_Y4.csv}\\
    \input{Tables2/table_rf_60h_baseline_Y4.csv}\\
    \input{Tables2/table_rf_52h_baseline_Y4.csv}\\
    \input{Tables2/table_rf_50h_baseline_Y4.csv}\\
    \input{Tables2/table_rf_34h_baseline_Y4.csv}\\
    \input{Tables2/table_rf_30h_baseline_Y4.csv}\\
    \input{Tables2/table_rf_22h_baseline_Y4.csv}\\
    \input{Tables2/table_rf_20h_baseline_Y4.csv}\\
    \bottomrule
	\end{tabularx}
\begin{tablenotes}[flushleft]
\item \emph{Notes.} ***, **, * refers to statistical significance at the 1, 5, and 10\% level. Robust standard errors for estimates of $\beta_Y$ in parentheses. The table shows estimated parameters $\alpha_Y$ and $\beta_Y$ (as presented in equation \ref{eqn:rf}) on secondary school track enrollment levels Low ($L_1$), Middle ($M_1$) and High ($H_1$), four years after the start of secondary education.
\end{tablenotes}
\end{threeparttable}
\end{center}
\end{singlespace}
\end{table}

Overall, the results suggests that for marginally assigned students, enrollment in a higher track in the first year increases the probability of higher track enrollment four years later. The strong positive effects at the $+ 2$ thresholds in particular may suggest deviations from predictions of our models. Specifically, our framework predicts that effects at the $+ 2$ thresholds should be weakly negative or close to zero. Certainly not strongly positive, relative to the amount of upgrading that appears to take place. These findings may therefore indicate a departure from the outcome-maximizing behavior we aim to assess in this paper. However, as noted earlier, we cannot ignore the role of parents (and students) in the decision-making process as they are potentially able to correct any systematic errors in the assignment behavior of teachers.


To integrate important qualitative aspects of the assignment process -- different kinds of shifts in the recommendation and first year enrollment -- as well as some features that are important for assessing the quality of the assignments -- both positive and negative effects --  we propose to move on with a flexible approach. The causal framework will make explicit that different kinds of treatments might take place at the same time at each threshold, and that these different treatments might have positive and/or negative effects. 
Our approach is able to harness all of these different causal effects and provides a systematic way of organizing them. The framework is general, but also allows for a straightforward way to simplify the structure by grouping the effects. Also, groupings that are not straightforward to operationalize using standard methodology, can be handled with ease within the context of our framework. 


%% file: Tables2/table_rf_60w_baseline_Y0.csv
havo + 2&0.867&-0.073*** (0.008)&0.111&-0.002 (0.007)&0.022&0.075*** (0.004)

%% file: Tables2/table_rf_52w_baseline_Y0.csv
vmbo-gt/havo + 2&0.709&-0.085*** (0.016)&0.279&0.040** (0.016)&0.012&0.045*** (0.006)

%% file: Tables2/table_rf_50w_baseline_Y0.csv
vmbo-gt + 2&0.898&-0.074*** (0.007)&0.087&0.014** (0.006)&0.015&0.059*** (0.004)

%% file: Tables2/table_rf_34w_baseline_Y0.csv
vmbo-kb/gt + 2&0.826&-0.172*** (0.024)&0.169&0.136*** (0.023)&0.004&0.036*** (0.007)

%% file: Tables2/table_rf_30w_baseline_Y0.csv
vmbo-kb + 2&0.981&-0.123*** (0.006)&0.007&0.035*** (0.004)&0.012&0.088*** (0.005)

%% file: Tables2/table_rf_22w_baseline_Y0.csv
vmbo-bb/kb + 2&0.936&-0.189*** (0.034)&0.057&0.160*** (0.032)&0.008&0.028** (0.014)

%% file: Tables2/table_rf_20w_baseline_Y0.csv
vmbo-bb + 2&0.901&-0.043*** (0.015)&0.044&-0.001 (0.011)&0.055&0.044*** (0.012)

%% file: Tables2/table_rf_61h_baseline_Y0.csv
havo/vwo + 1&0.974&-0.153*** (0.006)&0.026&0.153*** (0.006)&0.000&0.000 (0.000)

%% file: Tables2/table_rf_60h_baseline_Y0.csv
havo + 1&0.994&-0.068*** (0.003)&0.006&0.068*** (0.003)&0.001&0.000 (0.001)

%% file: Tables2/table_rf_52h_baseline_Y0.csv
vmbo-gt/havo + 1&0.977&-0.119*** (0.007)&0.021&0.124*** (0.007)&0.002&-0.005*** (0.002)

%% file: Tables2/table_rf_50h_baseline_Y0.csv
vmbo-gt + 1&0.994&-0.061*** (0.003)&0.005&0.060*** (0.003)&0.001&0.001 (0.001)

%% file: Tables2/table_rf_34h_baseline_Y0.csv
vmbo-kb/gt + 1&0.989&-0.118*** (0.010)&0.012&0.116*** (0.010)&-0.001&0.002* (0.001)

%% file: Tables2/table_rf_30h_baseline_Y0.csv
vmbo-kb + 1&0.998&-0.030*** (0.008)&0.001&0.020*** (0.007)&0.001&0.010*** (0.004)

%% file: Tables2/table_rf_22h_baseline_Y0.csv
vmbo-bb/kb + 1&0.979&-0.075*** (0.012)&0.018&0.089*** (0.011)&0.004&-0.014*** (0.005)

%% file: Tables2/table_rf_20h_baseline_Y0.csv
vmbo-bb + 1&1.000&-0.053*** (0.006)&-0.001&0.052*** (0.006)&0.002&0.001 (0.002)

%% file: Tables2/table_rf_60w_baseline_Y1.csv
havo + 2&0.248&-0.023** (0.010)&0.691&-0.027** (0.010)&0.062&0.050*** (0.006)

%% file: Tables2/table_rf_52w_baseline_Y1.csv
vmbo-gt/havo + 2&0.091&-0.003 (0.011)&0.572&-0.089*** (0.018)&0.337&0.092*** (0.017)

%% file: Tables2/table_rf_50w_baseline_Y1.csv
vmbo-gt + 2&0.507&-0.043*** (0.011)&0.452&-0.001 (0.011)&0.041&0.044*** (0.005)

%% file: Tables2/table_rf_34w_baseline_Y1.csv
vmbo-kb/gt + 2&0.196&-0.056** (0.023)&0.335&-0.056** (0.028)&0.469&0.112*** (0.029)

%% file: Tables2/table_rf_30w_baseline_Y1.csv
vmbo-kb + 2&0.636&-0.085*** (0.014)&0.283&0.004 (0.013)&0.081&0.080*** (0.009)

%% file: Tables2/table_rf_22w_baseline_Y1.csv
vmbo-bb/kb + 2&0.122&0.007 (0.043)&0.561&-0.039 (0.061)&0.317&0.032 (0.057)

%% file: Tables2/table_rf_20w_baseline_Y1.csv
vmbo-bb + 2&0.438&-0.014 (0.026)&0.467&0.002 (0.026)&0.095&0.011 (0.014)

%% file: Tables2/table_rf_61h_baseline_Y1.csv
havo/vwo + 1&0.037&0.006 (0.005)&0.824&-0.113*** (0.011)&0.139&0.107*** (0.010)

%% file: Tables2/table_rf_60h_baseline_Y1.csv
havo + 1&0.333&-0.023** (0.010)&0.656&0.021** (0.010)&0.011&0.002 (0.002)

%% file: Tables2/table_rf_52h_baseline_Y1.csv
vmbo-gt/havo + 1&0.157&-0.026** (0.012)&0.689&-0.038** (0.016)&0.154&0.064*** (0.012)

%% file: Tables2/table_rf_50h_baseline_Y1.csv
vmbo-gt + 1&0.600&-0.013 (0.011)&0.384&0.015 (0.011)&0.016&-0.001 (0.003)

%% file: Tables2/table_rf_34h_baseline_Y1.csv
vmbo-kb/gt + 1&0.244&-0.029 (0.025)&0.368&-0.020 (0.028)&0.388&0.049* (0.028)

%% file: Tables2/table_rf_30h_baseline_Y1.csv
vmbo-kb + 1&0.594&0.074* (0.039)&0.345&-0.050 (0.038)&0.061&-0.023 (0.018)

%% file: Tables2/table_rf_22h_baseline_Y1.csv
vmbo-bb/kb + 1&0.153&-0.019 (0.022)&0.608&0.014 (0.030)&0.239&0.005 (0.027)

%% file: Tables2/table_rf_20h_baseline_Y1.csv
vmbo-bb + 1&0.548&-0.043* (0.025)&0.426&0.026 (0.025)&0.026&0.017** (0.008)

%% file: Tables2/table_rf_60w_baseline_Y4.csv
havo + 2&0.130&-0.006 (0.008)&0.598&-0.036*** (0.011)&0.271&0.042*** (0.010)

%% file: Tables2/table_rf_52w_baseline_Y4.csv
vmbo-gt/havo + 2&0.028&-0.005 (0.006)&0.416&0.017 (0.018)&0.555&-0.012 (0.018)

%% file: Tables2/table_rf_50w_baseline_Y4.csv
vmbo-gt + 2&0.072&0.004 (0.006)&0.722&-0.031*** (0.010)&0.207&0.027*** (0.009)

%% file: Tables2/table_rf_34w_baseline_Y4.csv
vmbo-kb/gt + 2&0.014&0.010 (0.010)&0.319&-0.047* (0.027)&0.667&0.037 (0.028)

%% file: Tables2/table_rf_30w_baseline_Y4.csv
vmbo-kb + 2&0.081&-0.005 (0.009)&0.653&-0.072*** (0.014)&0.265&0.077*** (0.013)

%% file: Tables2/table_rf_22w_baseline_Y4.csv
vmbo-bb/kb + 2&-0.002&0.007 (0.005)&0.283&0.016 (0.057)&0.719&-0.023 (0.057)

%% file: Tables2/table_rf_20w_baseline_Y4.csv
vmbo-bb + 2&0.008&0.000 (0.004)&0.485&0.022 (0.026)&0.507&-0.021 (0.026)

%% file: Tables2/table_rf_61h_baseline_Y4.csv
havo/vwo + 1&0.077&-0.015** (0.007)&0.457&-0.024* (0.013)&0.465&0.038*** (0.013)

%% file: Tables2/table_rf_60h_baseline_Y4.csv
havo + 1&0.219&-0.011 (0.009)&0.651&-0.001 (0.010)&0.130&0.011 (0.007)

%% file: Tables2/table_rf_52h_baseline_Y4.csv
vmbo-gt/havo + 1&0.035&-0.004 (0.006)&0.545&-0.044*** (0.017)&0.421&0.048*** (0.016)

%% file: Tables2/table_rf_50h_baseline_Y4.csv
vmbo-gt + 1&0.111&-0.006 (0.007)&0.753&0.012 (0.010)&0.136&-0.006 (0.007)

%% file: Tables2/table_rf_34h_baseline_Y4.csv
vmbo-kb/gt + 1&0.046&0.000 (0.012)&0.340&0.001 (0.028)&0.614&-0.002 (0.028)

%% file: Tables2/table_rf_30h_baseline_Y4.csv
vmbo-kb + 1&0.146&-0.025 (0.029)&0.670&0.033 (0.038)&0.184&-0.008 (0.031)

%% file: Tables2/table_rf_22h_baseline_Y4.csv
vmbo-bb/kb + 1&0.001&0.002 (0.003)&0.236&-0.001 (0.028)&0.763&-0.001 (0.028)

%% file: Tables2/table_rf_20h_baseline_Y4.csv
vmbo-bb + 1&0.010&0.003 (0.006)&0.660&-0.009 (0.024)&0.330&0.006 (0.023)

%% file: Framework.tex
\section{Causal approach}\label{sec:framework}

Our previous results and discussion shows that a score above the cutoff may generate a cascade of effects on both the teacher track assignment, and first- and fourth-year track enrollment. This section introduces an extended IV framework that discipline these cascade of effects, which subsequently allows us to use the threshold effects to partially identify tracking effects and the quality of tacher track assignment. 

\subsection{A modified IV framework}

In a setting with one instrumental variable, we use the concept of principal strata introduced by \cite{FrangakisRubin2002} to modify the standard IV frameworks by \cite{imbens1994identification,angrist1995two} in three directions. First, our framework treats first-year track enrollment as multiple ordered treatments (Low, Mid, and High) in which we aim to identify their separate effects. Second, we similarly describe the fourth-year track enrollment as a multiple ordered outcome, and aim to identify the separate effects on each margin. Third, our framework includes the teacher track recommendation as a second treatment variable, next to first-year track enrollment.  

It will be convenient to summarize the three dummy variables of the track recommendation and track enrollment by the string variable $Et\in\{Lt,Mt,Ht\}$. For instance, $E_0=L_0$ when $L_0=1$, $E_1=M_1$ when $M_1=1$, and $E_4=H_4$ when $H_4=1$. A score above the cutoff may generate a cascade of effects. These effects are summarized by \Cref{fig:causalmodel}, where each arrow represents a possible effect among $\{Z, E_0, E_1, E_4\}$. First, a score above the cutoff prompts the teacher to reassess her track assignment, which in turn may lead to an upward revision of this assignment. \Cref{fig:causalmodel} refers to this as a teacher shift, where the variable $E_0(Z)$ is the potential track assignment of the teacher under each value of $Z$. For instance, an upgraded teacher assignment may look like this: $E_0(0)=L_0$ and $E_0(1)=M_0$, such that the teacher shifts the student from the low to the middle track when the student scores above the cutoff. 

\begin{figure}[t]
\centering
\caption{An overview of the possible effects generated by a score above the cutoff}
\label{fig:causalmodel}
\begin{tikzpicture}[shorten >=2pt,node distance=7.5cm,on grid,auto]
   \node[] (J) {$Z$};   
   \node[right=50mm of J] (K) {$E_0(z)$};   
   \node[right=50mm of K] (L) {$E_1(z,e_0)$};
   \node[right=50mm of L] (M)  {$E_4(e_1)$};
   \path[->]
    (J) edge [left]  node [above] {\emph{teacher shift}} (K) 
    (J) edge [bend left]  node[above] {\emph{parental shift}} (L) 
    (K) edge [left]  node [above] {\shortstack{\emph{converted} \\ \emph{teacher shift}}} (L)
    (L) edge [left]  node [above] {\emph{tracking effect} } (M) ;
\end{tikzpicture}
\caption*{ \footnotesize
\emph{Notes.} The teacher assignment $E_0(z)$ can only be affected by $Z$, first-year track enrollment $E_1(z,e_0)$ can be affected by both $Z$ and $E_0$, and fourth-year track enrollment $E_4(e_1)$ can only be affected by $e_1$.}
\end{figure}

Second, first-year track enrollment may be affected in two ways: Either the upwardly revised track assignment is converted into a higher first-year track enrollment, or as score above the cutoff may directly lead to a higher first-year track enrollment. \Cref{fig:causalmodel} refers to the first mechanism as a converted teacher shift, and the second as a parental shift, where either the parent uses a score above the cutoff to pressure the high school into increasing first-year enrollment, or the high school acts on its own, but with the parent's agreement. To capture these two pathways, potential first-year track enrollment is a function of both variables, $E_1(z,e_0)$. A converted teacher shift is described by a change in $E_1$ when both $Z$ and $E_0$ change, whereas a parental shift is described by a change in $E_1$ when only $Z$ changes and keeping $E_0$ fixed. 

Finally, fourth-year track enrollment is only affected by first-year track enrollment. \Cref{fig:causalmodel} refers to this as the tracking effect, where potential fourth-year track enrollment is as a function of first-year track enrollment only, $E_4(e_1)$. This single pathway implies that an upwards revision by the teacher due to a score above the cutoff, without a change in first-year track enrollment, cannot change fourth-year track enrollment. 

We formalize the description and identification of the causal model in \Cref{fig:causalmodel} with the following set of assumptions:

\begin{assumption}[Assumptions causal framework]\label{ass:iv} \
\renewcommand{\theenumi}{\alph{enumi}}
\begin{enumerate}
    \item (Continuity) $E_{0}(z)$, $E_{1}(z,e_0)$, and $E_{4}(e_1)$ are continuous in $S$ at $S=0$ $\forall$  $z,e_{0},e_{1}$.
    \item (Monotonicity $E_0$) $L_0(1)\leq L_0(0)$ and $H_0(1)\geq H_0(0)$ $\forall$  students, \\
    (Monotonicity $E_1$) $L_1(1,E_0(1))\leq L_1(0,E_0(0))$ and $H_1(1,E_0(1))\geq H_1(0,E_0(0))$ $\forall$ students.
    \item (Exclusion) $E_4(z,e_0,e_1)=E_4(e_1)$ $\forall$  $z,e_{0},e_{1}$.
    \end{enumerate} 
\end{assumption}
Assumption \ref{ass:iv}a is the standard continuity assumption required to identify the threshold effects: Students just to the left and right of the test score cutoffs have similar potential outcomes, and hence are similar on ex-ante characteristics. Similar to the standard monotonicity assumption, Assumption \ref{ass:iv}b requires that scoring above the cutoff can only shift students towards a higher teacher track assignment and towards a higher first-year track enrollment. However, in our framework with three tracks, this implies that we allow for three positive assignment and first-year enrollment shifts: from Low to Middle, from Low to High, and from Middle to High. Importantly, Assumption \ref{ass:iv}.b allows for both shifts away from and towards the Middle track. Assumption \ref{ass:iv}c imposes the exclusion restriction that scoring above the cutoff only has an effect on fourth-year track enrollment if it also affects first-year track enrollment. 

\subsection{Partial identification}

We aim to use our causal framework the identify tracking effects and test for the quality of teacher track assignments. Although the framework restricts the effects generated by a score above the cutoff, point identification of all the remaining effects is generally impossible with just a single instrument or without additional assumptions. Therefore, we will resort to a partial identification approach. 

We explain the intuition behind our strategy through a discussion on the first step in the framework: The teacher track revision. Under \Cref{ass:iv}, the threshold effects on the three revision dummies contain the proportions of students who shift in a manner that involves each respective track:
\begin{align}
    \beta_{L_0}=& \underset{s \to^+ 0}{\lim}\mathbb{E}[L_{0}|S=s]-\underset{s \to^-0}{\lim}\mathbb{E}[L_{0}|S=s]=\mathbb{E}[L_{0}(1)-L_{0}(0)|S=0] \\ \notag
    =&  -\mathbb{P}[L_0(1)-L_0(0)=-1|S=0] \\ \notag
    =& -\mathbb{P}[E_0(1)\neq L_0, E_0(0)=L_0|S=0] \\ \notag
    =& -\mathbb{P}[E_0(1)=M_0, E_0(0)=L_0|S=0]-\mathbb{P}[E_0(1)=H_0, E_0(0)=L_0|S=0]\\ \notag
    =& -\mathbb{P}[L_0 \rightarrow M_0]-\mathbb{P}[L_0 \rightarrow H_0],\\
    \beta_{M_0}=& \mathbb{P}[L_0 \rightarrow M_0]-\mathbb{P}[M_0 \rightarrow H_0],\\
    \beta_{H_0}=& \mathbb{P}[L_0 \rightarrow H_0]+\mathbb{P}[M_0 \rightarrow H_0].
\end{align}
Our mostly negative estimates on $L_0$ point identifies the total proportion of students that shift way from a Low assignment, but from this we cannot point identify the proportion that shifts from Low towards Middle versus from Low towards High. Similarly, our mostly negative estimates on $M_0$ imply that the proportion of students that shift away from the Middle assignment is larger than the proportion that shift towards it, but point identification of these separate proportions is generally impossible. Finally, our mostly positive estimates on $H_0$ point identifies the total proportion of students that shift towards the High assignment, but again we cannot point identify where they come from. 

Additional information on these three proportions may be contained in the control means (just below the cutoff) for each of the three teacher track assignment dummies. Under \Cref{ass:iv}, the control mean contains the proportion of students who receive that track assignment when they score below the cutoff:
\begin{align}
    \alpha_{L_0}=&  \underset{s \to^-0}{\lim}\mathbb{E}[L_{0}|S=s] = \mathbb{E}[L_{0}(0)|S=0] =\mathbb{P}[ L_0(0)=1|S=0] \\ \notag
    =& \mathbb{P}[ E_0(0)=L_0|S=0]\\ \notag
    =&  \mathbb{P}[L_0 \rightarrow L_0] + \mathbb{P}[L_0 \rightarrow M_0]+\mathbb{P}[L_0 \rightarrow H_0],\\
    \alpha_{M_0}=&  \mathbb{P}[M_0 \rightarrow M_0] + \mathbb{P}[M_0 \rightarrow H_0],\\
    \alpha_{H_0}=& \mathbb{P}[H_0 \rightarrow H_0].
\end{align}
On top of the proportions for the three upward shifts, the control means also contain three proportion of students that do not shift at the threshold. For instance, on top of two proportion of shifts from Low to Middle and Low to High, the control mean on the Low assignment also contains the proportion of individuals that always receive Low. Besides information on the proportion of these non-shifters, the control means may provide additional information on the proportion of three shifts. For instance, if the control mean for Middle assignment is zero, we know that the proportion of shifts from Middle to High is zero, which allows us to point identify the three proportions in combination with the threshold effects above.

We can combine the control means and threshold effects to generate the treatment means, which contain the proportion of students who receive that track assignment  when they score above the cutoff:
\begin{align}
    \alpha_{L_0}+\beta_{L_0}=& \mathbb{P}[L_0 \rightarrow L_0] ,\\ 
    \alpha_{M_0}+\beta_{M_0}=& \mathbb{P}[M_0 \rightarrow M_0] + \mathbb{P}[L_0 \rightarrow M_0],\\ 
    \alpha_{H_0}+\beta_{H_0}=&  \mathbb{P}[H_0 \rightarrow H_0] + \mathbb{P}[L_0 \rightarrow H_0]+  \mathbb{P}[M_0 \rightarrow H_0].
\end{align}

In general, the control and treatment means are observed probability distributions that contain all the available information about the unobserved proportions of potential outcomes. Beyond this, no additional information can be retrieved about the potential outcomes. It is informative to organize these probability distributions in a polytope, which is defined as a matrix of non-negative numbers whose row and column sums equal the corresponding margin \citep{de2009graphs}. The polytope for the teacher revision can be written as: 
\begin{equation}\label{eqn:polytope3}
\left[
\begin{array}{c|ccc}
& \alpha_{L_0} + \beta_{L_0} & \alpha_{M_0} + \beta_{M_0} & \alpha_{H_0} + \beta_{H_0} \\
\hline
\alpha_{L_0}& \mathbb{P}[L_0 \rightarrow L_0] & \mathbb{P}[L_0 \rightarrow M_0] &  \mathbb{P}[L_0 \rightarrow H_0] \\
\alpha_{M_0} & & \mathbb{P}[M_0 \rightarrow M_0] &  \mathbb{P}[M_0 \rightarrow H_0]   \\
\alpha_{H_0} & & &  \mathbb{P}[H_0 \rightarrow H_0] 
\end{array}
\right]
\end{equation}
Each row and column reflects, respectively, the control and treatment mean of the corresponding assignment dummy.  We will refer to the three sifts and three non-shifts as the six student ``types''. The entry in each row-column combination contains the student type that is present in the control mean of that row and in the treatment mean of that column. This implies that each non-diagonal entry contains the type that shifts away from the outcome in the row towards the outcome in the column due to a score above the cutoff, whereas each diagonal entry contains the type of non-shifters corresponding to the outcome in the row (and column). Empty entries are shifts ruled out by \Cref{ass:iv}. In this case, all empty entries reflect the monotonicity assumption \ref{ass:iv}b.

We subsequently aim to retrieve solutions to these six proportions of types. There are potentially many solutions, but our solutions of interest are the ones that minimize and maximize each proportion, or combinations thereof, while satisfying the row-sum and column-sum restrictions. Hence, we aim to find the upper and lower bound for each proportion of types subject to the equality constraint defined by the polytope. This can be expressed as a standard linear programming problem, where we aim to find a $6 \times 1$ vector $x$, containing the proportion of student types, as follows:
\begin{align}\label{eqn:lp}
\min_{x} / \max_{x} \quad  g^{\top}x \quad \text{subject to} \quad Ax = b, \quad x\geq 0,
\end{align}
where $g$ is a  $6 \times 1$ vector of zeros and ones describing the proportions to be minimized or maximized, $A$ is $6 \times 6$  matrix of zeros and ones describing the relationship between the six student types and the control and treatment means, and $b$ is a $6 \times 1$ vector containing the control and treatment means. For the polytope in \eqref{eqn:polytope3}, the equality $Ax=b$ looks as follows:
\begin{equation*}
\begin{bmatrix}
1 & 1 & 1 & 0 & 0 & 0 \\
0 & 0 & 0 & 1 & 1 & 0 \\
0 & 0 & 0 & 0 & 0 & 1 \\
1 & 0 & 0 & 0 & 0 & 0 \\
0 & 1 & 0 & 1 & 0 & 0 \\
0 & 0 & 1 & 0 & 1 & 1 \\
\end{bmatrix}
\begin{bmatrix}
\mathbb{P}[L_0 \rightarrow L_0] \\ 
\mathbb{P}[L_0 \rightarrow M_0] \\ 
\mathbb{P}[L_0 \rightarrow H_0] \\ 
\mathbb{P}[M_0 \rightarrow M_0] \\
\mathbb{P}[M_0 \rightarrow H_0] \\ 
\mathbb{P}[H_0 \rightarrow H_0]
\end{bmatrix}
=
\begin{bmatrix}
\alpha_{L_0} \\ 
\alpha_{M_0}  \\ 
\alpha_{H_0}  \\ 
\alpha_{L_0} + \beta_{L_0} \\ 
\alpha_{M_0} + \beta_{M_0} \\ 
\alpha_{H_0} + \beta_{H_0} 
\end{bmatrix}
.
\end{equation*}
We can now, for instance, find the lower and upper bound for the assignment shift from Low to Middle by setting the second entry in $g$ equal to one, and the other entries to zero. We can repeat this for any other proportions, or combinations thereof. Note that, as the lower and upper bound hold with equality, the bounds are sharp by construction: They are the largest lower and smallest upper bound given the assumptions and what can be identified from the data.


\subsection{Testing assumptions and finding solutions}

It is possible that no vector $x$ exists as a solution to the linear programming problem in \eqref{eqn:lp}. Besides sampling uncertainty in the estimates for the control and treatment mean, this suggests that  IV \cref{ass:iv} is rejected. Consider, for instance that $\alpha_{L_0}+\beta_{L_0}>\alpha_{L_0}$: The treatment mean on the Low track is larger than the control mean on the low track. According to the polytope in \eqref{eqn:polytope3} this cannot happen, as the treatment mean only contains the proportion of non-shifters in Low, whereas the control mean contains contains the same proportion of non-shifters in Low and two proportions that shift away from Low. This would suggest that our monotonicity assumption is violated, and that students may also shift towards Low (from Middle or High), such that these two empty entries in the first column in \eqref{eqn:polytope3} would be filled with $\mathbb{P}[M_0 \rightarrow L_0]$ and $\mathbb{P}[H_0 \rightarrow L_0]$.

We can use our linear programming procedure to develop a (heuristic) test for the IV assumptions. We do this by introducing 12 additional slack parameters in the $12 \times 1$ vector $s$, two parameters for each control and treatment mean. We subsequently augment the linear programming problem as follows:
\begin{align}\label{eqn:slack}
\min_{x,s} \quad 
\begin{bmatrix}
    g^{\top} & g_s^{\top}
\end{bmatrix}
\begin{bmatrix}
    x \\ 
    s
\end{bmatrix}
\quad 
\text{subject to} 
\quad 
\begin{bmatrix}
    A & I & -I
\end{bmatrix}
\begin{bmatrix}
    x \\
    s
\end{bmatrix}
= b, \quad x\geq 0, \quad s\geq 0,
\end{align} 
where $g_s$ is a  $12 \times 1$ vector of ones describing the proportions of slack to be minimized and $I$ is a $6 \times 6$ identity matrix. It is immediate there always exists a vector $[x, s]$ as a solution to the problem in \eqref{eqn:slack}. However, does there exist a solution with sufficiently small slack? If not, this suggests that \Cref{ass:iv} is violated. 

To test our IV assumptions, we set all entries in $g$ to zero and all entries in $g_s$ to one, which minimizes the total proportion of slack. If total slack is close to zero, this suggests suggests that we cannot reject IV assumption \ref{ass:iv}. Besides this heuristic test, we use the slack variables to guarantee consistent solutions for the vector of student types $x$. In particular, we store the estimated slack in the $12\times 1$ vector $b_s$ and augment the linear programming problem in \eqref{eqn:lp} as follows:
\begin{align}\label{eqn:lpslack}
\min_{x} / \max_{x} \quad
\begin{bmatrix}
    g^{\top} & g_s^{\top}
\end{bmatrix}
\begin{bmatrix}
    x \\ 
    s
\end{bmatrix}
\quad 
\text{subject to} 
\quad 
\begin{bmatrix}
    A & I & -I\\
    0 & I & 0\\
    0 & 0 & I\\
\end{bmatrix}
\begin{bmatrix}
    x \\
    s
\end{bmatrix}
= 
\begin{bmatrix}
    b\\
    b_s
\end{bmatrix}
, \quad x\geq 0.
\end{align} 
We set all entries in $g_s$ to zero, and the entries in $g$ to one that correspond to the student types for whom we aim to find the lower and upper bound. This formulation ensures that the 12 additional slack parameters are fixed at their values $b_s$ estimated to initially test our assumptions, such that solutions exists to the student types.  Whether these solutions can be interpreted as meaningful lower and upper bounds depends on the degree of slack.

\subsection{Applying the approach to tracking effects}

This section combines the causal model with the partial identification approach to discuss the identification tracking effects: The effect of first-year track enrollment $E_1$ on fourth-year track enrollment $E_4$. 
Recall that \Cref{fig:causalmodel} shows that first-year track enrollment depends on $Z$ both indirectly through the teacher shift and directly through the parental shift. Our pursuit of tracking effects, however, does not require us to disentangle these two mechanisms, and we simplify potential first-year track enrollment by writing $E_1(Z,E_0(Z))=E_1(Z)$. Hence, this section does not make use of the potential teacher revision $E_0$.\footnote{Our causal model now mimics an IV setting, where the score above the cutoff can be used as an instrument for first-year track enrollment, to identify tracking effects on fourth-year track enrollment. Note that our framework does not allow for the identification of the teacher track assignment effects on fourth-year track enrollment. The reason is that a score above the cutoff also affects fourth-year track enrollment without affecting teacher track assignment, via first-year track enrollment. } 

Similar to the discussion for the teacher assignment, \Cref{ass:iv} allows a score above the cutoff to generate three types of positive shifts in first-year track enrollment: Low to Middle, Low to High, and Middle to High. The assumptions, however, do not restrict the tracking effects of first-year track enrollment on fourth-year track enrollment. Hence, each of these three first-year enrollment shifts may generate three types of positive tracking effects, three types of negative tracking effects, and three types of null effects on fourth-year track enrollment. For instance, a first-year enrollment shift from Low to Middle may generate a positive fourth-year tracking effect from Low to Middle, a negative tracking effect from Middle to Low, or no tracking effect as the student always ends up on the Low track. This implies we have a total of $3\times 9=27$ potential shifts that may be generated by a score above the cutoff. 

On top of these 27 shifts, there are also 9 non-shifts. Similar to the discussion for the teacher track assignment, a student may always enroll in the Low, Middle, or High track in the first year, despite a score above the cutoff. And for each of these three first year track enrollments, we may observe the student in the Low, Middle, or High track in fourth year. Hence, this generates $3\times 3=9$ potential non-shifts. In total, we thus have $27+9=36$ student types when analyzing tracking effects.

To test for tracking effects, it will be useful to introduce notation for the 36 types and initially categorize  them into 6 broad categories:
\begin{itemize}
    \item \textbf{Trapped in Track} $TT$: Higher track enrollment in first year implies higher track enrollment in fourth year.
    \begin{itemize}
        \item $TT=\{ 
        TT_{LM_{1}}^{LM_{4}},
        TT_{LM_{1}}^{LH_{4}},
        TT_{LM_{1}}^{MH_{4}},
        TT_{LH_{1}}^{LM_{4}},
        TT_{LH_{1}}^{LH_{4}},
        TT_{LH_{1}}^{MH_{4}},
        TT_{MH_{1}}^{LM_{4}},
        TT_{MH_{1}}^{LH_{4}},
        TT_{MH_{1}}^{MH_{4}}\}$
    \end{itemize}
    \item \textbf{Slow Starters} $SS$: Higher track enrollment in first year implies lower track enrollment in fourth year. 
    \begin{itemize}
        \item $SS=\{ 
        SS_{LM_{1}}^{ML_{4}},
        SS_{LM_{1}}^{HL_{4}},
        SS_{LM_{1}}^{HM_{4}},
        SS_{LH_{1}}^{ML_{4}},
        SS_{LH_{1}}^{HL_{4}},
        SS_{LH_{1}}^{HM_{4}},
        SS_{MH_{1}}^{ML_{4}},
        SS_{MH_{1}}^{HL_{4}},
        SS_{MH_{1}}^{HM_{4}}\}$
    \end{itemize}   
    \item \textbf{Always Low} $AL$: Higher track enrollment in first year does not affect track enrollment in fourth year, with fourth-year enrollment always in the low track.
    \begin{itemize}
        \item $AL=\{ 
        AL_{LM_{1}},
        AL_{LH_{1}},
        AL_{MH_{1}}\}$    
    \end{itemize}
    \item \textbf{Always Middle} $AM$: Higher track enrollment in first year does not affect track enrollment in fourth year, with fourth-year enrollment always in the middle track.
    \begin{itemize}
        \item $AM=\{ 
        AM_{LM_{1}},
        AM_{LH_{1}},
        AM_{MH_{1}}\}$    
    \end{itemize}
    \item \textbf{Always High} $AH$: Higher track enrollment in first year does not affect track enrollment in fourth year, with fourth-year enrollment always in the high track.
    \begin{itemize}
        \item $AH=\{ 
        AH_{LM_{1}},
        AH_{LH_{1}},
        AH_{MH_{1}}\}$    
    \end{itemize}
    \item \textbf{Non-Shifters} $NS$: Track enrollment in first year is not affected by a score above the cutoff, with first-year enrollment always in the low, middle, or high track.
    \begin{itemize}
        \item $NS=\{ 
        L_{L_{1}},
        L_{M_{1}},
        L_{H_{1}},
        M_{L_{1}},
        M_{M_{1}},
        M_{H_{1}},        
        H_{L_{1}},
        H_{M_{1}},
        H_{H_{1}} \}$    
    \end{itemize}  
\end{itemize}
The 27 shifters are bundled in the first 5 categories, and the 9 non-shifters in the last category. The 5 categories of the shifters are based upon how the first-year enrollment shift affects the fourth-year enrollment: positively, negatively, or not at all. The first-year enrollment shift is in the subscript and, for the two broad types affected by this first-year shift, the fourth-year enrollment shift is in the superscript. For instance, the positively affected Trapped in Tracker $TT_{LM_1}^{LH_4}$ shifts from Low to Middle in first year due to a score above the cutoff and, as a result, shifts from Low to High in fourth year. In contrast, the negatively affected Slow Starter $SS_{LM_1}^{HL_4}$ experiences the same first-year enrollment shift, but as a result shifts from High to Low after four years. The students unaffected by the shift from Low to Middle in first year may always end up in Low ($AL_{LM}$), Middle ($AH_{LM}$), or High ($AH_{LM}$) after four years. The final sixth category of Non-Shifters are characterized by their single level of first-year track enrollment in the subscript. For instance, the students always starting on the Low track in first year, may either end up in Low ($L_{L_1}$), Middle ($M_{L_1}$), or High ($H_{L_1}$).
 
Similar to before, we can build a polytope that captures the relationship between the proportions of types and the control and treatment means. Recall that each (off-diagonal) entry of the polytope contains a single type that shifts away from the control outcome towards the treatment outcome. To make sure each entry contains one type only, we need to take the control and treatment means of the interaction between first- and fourth-year enrollment, as each type is defined by both first- and fourth-year enrollment. As we have three first- and fourth-year enrollment dummies, we have $3\times 3=9$ interacted outcome variables. This guarantees that we look at the complete probability distribution of our data, such that not more cannot be learned from the observed distributions about the unobserved proportions of types.

For each of the nine outcome variables, we can connect the control and treatment mean to the 36 proportions of student types under \Cref{ass:iv}. For instance, the interaction between the Low track enrollment dummies in the first and fourth year contain the following proportions of types:
\begin{align}
    \alpha_{L_1 L_4}=&  \underset{s \to^-0}{\lim}\mathbb{E}[L_1 \times L_4|S=s] = \mathbb{E}[L_1(0)L_4(L_1)|S=0] =\mathbb{P}[L_1(0)L_4(L_1)=1|S=0] \\ \notag
    =& \mathbb{P}[ E_1(0)=L_1, E_4(L_1)=L_1 |S=0]\\ \notag
    =&  \mathbb{P}[L_{\scriptscriptstyle L_1}] +  \mathbb{P}[AL_{\scriptscriptstyle LM_1}] +  \mathbb{P}[TT_{\scriptscriptstyle LM_1}^{\scriptscriptstyle LM_4}] +  \mathbb{P}[TT_{\scriptscriptstyle LM_1}^{\scriptscriptstyle LH_4}] +  \mathbb{P}[AL_{\scriptscriptstyle LH_1}] +  \mathbb{P}[TT_{\scriptscriptstyle LH_1}^{\scriptscriptstyle LM_4}] +  \mathbb{P}[TT_{\scriptscriptstyle LH_1}^{\scriptscriptstyle LH_4}] ,\\
    \alpha_{L_1 L_4} + \beta_{L_1 L_4}= &  \underset{s \to^+0}{\lim}\mathbb{E}[L_1 \times L_4|S=s] = \mathbb{E}[L_1(1)L_4(L_1)|S=0] =\mathbb{P}[L_1(1)L_4(L_1)=1|S=0] \\ \notag
    =& \mathbb{P}[ E_1(1)=L_1, E_4(L_1)=L_1 |S=0]\\ \notag
    =&  \mathbb{P}[L_{\scriptscriptstyle L_1}] .
\end{align}
The control mean contains the seven student types that are in the Low track in year one and four when they score below the cutoff, whereas the treatment mean contains one type that is in the Low track in both years when they score above the cutoff. Repeating this for the other eight interacted variables, allows us to make the following polytope:
\begin{equation}\label{eqn:polytope33}
\left[
\begin{array}{c|ccccccccc}
& L_1 L_4 & L_1 M_4 & L_1 H_4 & M_1 L_4 & M_1 M_4 & M_1 H_4 & H_1 L_4 & H_1 M_4 & H_1 H_4 \\
\hline 
\noalign{\vskip 0.4em}
L_1 L_4	&	L_{\scriptscriptstyle L_{1}}	&		&		&	AL_{\scriptscriptstyle LM_{1}}	&	TT_{\scriptscriptstyle LM_{1}}^{\scriptscriptstyle LM_{4}}  &	TT_{\scriptscriptstyle LM_{1}}^{\scriptscriptstyle LH_{4}}	&	AL_{\scriptscriptstyle LH_{1}}	&	TT_{\scriptscriptstyle LH_{1}}^{\scriptscriptstyle LM_{4}}	&	TT_{\scriptscriptstyle LH_{1}}^{\scriptscriptstyle LH_{4}}	\\ [.4em]
L_1 M_4		&		&	M_{\scriptscriptstyle L_{1}}	&		&	SS_{\scriptscriptstyle LM_{1}}^{\scriptscriptstyle ML_{4}}	&	AM_{\scriptscriptstyle LM_{1}}	&	TT_{\scriptscriptstyle LM_{1}}^{\scriptscriptstyle MH_{4}}	&	SS_{\scriptscriptstyle LH_{1}}^{\scriptscriptstyle ML_{4}}	&	AM_{\scriptscriptstyle LH_{1}}	&	TT_{\scriptscriptstyle LH_{1}}^{\scriptscriptstyle MH_{4}} \\ [.4em]
L_1 H_4		&		&		&	H_{\scriptscriptstyle L_{1}}	&	SS_{\scriptscriptstyle LM_{1}}^{\scriptscriptstyle HL_{4}}	&	SS_{\scriptscriptstyle LM_{1}}^{\scriptscriptstyle HM_{4}}	&	AH_{\scriptscriptstyle LM_{1}}	&	SS_{\scriptscriptstyle LH_{1}}^{\scriptscriptstyle HL_{4}}	&	SS_{\scriptscriptstyle LH_{1}}^{\scriptscriptstyle HM_{4}} & AH_{\scriptscriptstyle LH_{1}} \\ [.4em]
M_1 L_4		&		&		&		&	L_{\scriptscriptstyle M_{1}}	&		&		&	        AL_{\scriptscriptstyle MH_{1}}  &	TT_{\scriptscriptstyle MH_{1}}^{\scriptscriptstyle LM_{4}}	&	TT_{\scriptscriptstyle MH_{1}}^{\scriptscriptstyle LH_{4}}	\\ [.4em]
M_1 M_4		&		&		&		&		&	M_{\scriptscriptstyle M_{1}}	&		&	SS_{\scriptscriptstyle MH_{1}}^{\scriptscriptstyle ML_{4}}	&	AM_{\scriptscriptstyle MH_{1}}	&	TT_{\scriptscriptstyle MH_{1}}^{\scriptscriptstyle MH_{4}}	\\ [.4em]
M_1 H_4		&		&		&		&		&		&	H_{\scriptscriptstyle M_{1}}	&	SS_{\scriptscriptstyle MH_{1}}^{\scriptscriptstyle HL_{4}}	&	SS_{\scriptscriptstyle MH_{1}}^{\scriptscriptstyle HM_{4}}	&	AH_{\scriptscriptstyle MH_{1}}	\\ [.4em]
H_1 L_4		&		&		&		&		&		&		&	L_{\scriptscriptstyle H_{1}}	&		&		\\ [.4em]
H_1  M_4		&		&		&		&		&		&		&		&	M_{\scriptscriptstyle H_{1}}	&		\\ [.4em]
H_1 H_4		&		&		&		&		&		&		&		&		&	H_{\scriptscriptstyle H_{1}}  
\end{array}
\right]
\end{equation}
Similar to \eqref{eqn:polytope3}, the rows reflect the control mean of the interacted outcome and the columns reflect the treatment mean of the interacted outcome. An (off-diagonal) entry contains the proportion of students that shift out of the interacted outcome in the row, and into the interacted outcome in the column, due to a score above the cutoff. Empty entries are student types ruled out by \Cref{ass:iv}. Note that each empty entry can be linked to one specific IV assumption. In particular, all empty entries above the diagonal are ruled out by the exclusion restriction in \Cref{ass:iv}c, and most empty entries below the diagonal are ruled out by monotonicity in \Cref{ass:iv}b.

The procedure now follows similarly as before. We can find the solutions of the proportions of types through linear programming. In this case, the $A$ matrix is of dimension $18\times 36$, and the vectors $x$, $b$, and $s$ are of dimensions $36\times 1$, $18\times 1$, and $36 \times 1$, respectively. 
It is important to stress again that we can partially identify any combination of student types, such as the total proportion of Trapped in Trackers by setting the 9 types in $TT$  equal to one in the $g$-vector, or the total proportions of students that experience tracking effects, by setting the 18 types in $TT$ and $SS$  equal to one in the $g$-vector. This allows us to test for combinations of effects that is in general not possible with previous approaches. This is in our view the main useful feature behind our approach. 

\subsection{Applying the approach to test quality of track assignment}

The previous section analyzed tracking effects and clarified there were three potential shifts in first-year track enrollment: Low to Mid, Low to High, and Middle to High. \Cref{fig:causalmodel} shows that any of these three shifts can either be the result of a (converted) teacher shift, or a parental shift. This section aims to extend our framework to separate between these two mechanisms such that we can analyze the quality of track assignment. For instance, in case we find tracking effects generated by the three shifts in first year, do these shifts originate from a (converted) teacher shift, the parental shift, or both? The answer to this question is crucial for the analyses on the quality of track assignment.

We therefore extend our framework to include the teacher track assignment $E_0$. Recall that, similar to the first-year track assignment, the teacher track assignment also has six student types. There are three shifts (Low to Middle, Low to High, and Middle to High) and three non-shifts (Low, Middle, and High). \Cref{fig:causalmodel} clarifies that a shift or non-shift in the teacher assignment  can been seen as the mechanism through which a score above the cutoff affects first-year track enrollment, but otherwise does not restrict the potential shifts and non-shifts in first-year track enrollment. 

This implies that for each of the 36 student types discussed for the analyses of tracking effects, there are six potential versions depending on their teacher assignment shift (or non-shift). For instance, consider the type $TT_{LM_1}^{LM_1}$, who is positively affected by the first-year enrollment shift from Low to Middle. There are six potential version of this Trapped in Tracker: Those that start with one of three potential shifts or non-shifts in the teacher assignment. This is similar for the remaining 35 students types discussed above. Hence, the inclusion of the teacher assignment implies we have $6\times 36=216$ types. The unobserved proportions of these types are captured by the control and treatment means of the interacted outcome variables. As we have three dummies for the teacher assignment, and first- and fourth-year enrollment, we have  $3\times 3 \times 3 =27$ interacted outcome variables. 

The linear programming procedure now follows similarly as before. In this case, the $A$ matrix is of dimensions $54\times 216$, the vectors $x$, $b$, $s$, and $b_s$ are of dimensions $216\times 1$, $54\times 1$, $108\times 1$, and $108\times 1$ respectively. As we can partially identify any combination of student types, we can use this version of the problem to also identify the tracking effects above, or any other combination of effects. For instance, to identify the total proportion of Trapped in Trackers we can set the the $6\times 9=54$ types in $TT$  equal to one in the vector $g$. 

To implement our procedure, we estimate \eqref{eqn:rf} with the 27 interaction terms as outcome variables to retrieve the control and treatment means for the vector $b$.  We round our estimates to six decimal places. We use the predictor-corrector primal-dual method by \cite{mehrotra1992implementation} to find our solutions. In a first step, we estimate the slack variables in $s$ and store them in the vector $b_s$. In a second step, we estimate the proportion of types in $x$ while keeping slack fixed. We can implement this second step for any combination of types that is of interest by altering the vector $g$. To obtain standard errors for the proportion of types, we repeat this procedure for 1000 bootstrapped samples.

Our approach to use slack as a test for the IV assumptions relates to the specification test developed by \cite{kitagawa2015test}, which essentially uses interactions between binarized treatment and outcome variables to test necessary conditions of IV validity obtained by \cite{balke1997bounds,imbens1997estimating,heckman2005structural}. These necessary conditions amount to similar observations discussed above: The treatment mean on $L_0\times L_1\times L_4$ cannot be larger than the control mean under \Cref{ass:iv}. We extend this to a procedure that simultaneously can test all restrictions on the treatment and outcome variables implied by the IV assumption. Developing the procedure into a formal test would require knowledge of the distribution of total slack under the null hypothesis of no violation of the IV assumptions. Moreover, as each empty cell in the polytope can be linked to one specific IV assumption, a formal test could separately assess the validity of the monotonicity assumption and exclusion restriction, instead of jointly as in \cite{kitagawa2015test}. We consider the development of a formal test beyond the scope of this draft.

%% file: Results.tex
\section{Results}\label{sec:results}

We use the RD model in \eqref{eqn:rf} to estimate, for each initial recommendation, the control and treatment mean for the set of 27 interacted variables between $E_0$, $E_1$ and $E_4$. After relying on the specific IV-type assumptions, we still have a system that is underidentified. In fact, in our general model we have 216 causal effects, or principal strata, and only 54 equations. The purpose of these strata is also not to also obtain point identification for each of these 216 types. This framework is merely a starting point to group these 216 types in ways that are relevant for our purpose. These groupings of types are often also not point identified, only bounded. These bounds however, as we show, can be very informative. 

\subsection{Tracking effects}\label{sec:tracking_effects}

In this section, we use our framework to investigate the tracking effects for marginally assigned students. The question of whether assignment is optimal presupposes that tracking effects exist. Without such effects at the margin, any assignment should be considered optimal by default. 

Tracking effects arise when a student's fourth year track enrollment is affected by track enrollment in the first year. To provide the right context, we first examine the effects on first year track enrollment. As in previous sections, we consider the three track levels — $L_1$, $M_1$, and $H_1$ — and the associated shifts: $L_1 \rightarrow M_1$, $M_1 \rightarrow H_1$, and $L_1 \rightarrow H_1$. In Table \ref{tab:first_stage}, we apply our causal approach to present estimates for these three types of shifts that might occur in the data. 
In the first two columns of Table \ref{tab:first_stage} we show bounds on the fraction of students who have experienced any shift in first year enrollment.

\begin{table}[!h]
\begin{singlespace}
    \begin{center}
    \begin{threeparttable}
    \caption{\label{tab:first_stage} Track shifts}
	\begin{tabularx}{1\textwidth}{@{\extracolsep{\fill}} lcccccccc}
		\toprule
         &\multicolumn{2}{c}{any shift}&\multicolumn{2}{c}{$L_1\rightarrow M_1$}&\multicolumn{2}{c}{$L_1\rightarrow H_1$}&\multicolumn{2}{c}{$M_1\rightarrow H_1$} \\ 
         \cmidrule(lr){2-3}\cmidrule(lr){4-9}
        &LB&UB&LB&UB&LB&UB&LB&UB\\    
            \cmidrule(lr){2-9}
            &(1)&(2)&(3)&(4)&(5)&(6)&(7)&(8)\\
		\cmidrule(lr){2-9}
    \input{Tables2/table_bline_e4_2_60w.csv}\\
    \input{Tables2/table_bline_e4_2_52w.csv}\\
    \input{Tables2/table_bline_e4_2_50w.csv}\\
    \input{Tables2/table_bline_e4_2_34w.csv}\\
    \input{Tables2/table_bline_e4_2_30w.csv}\\
    \input{Tables2/table_bline_e4_2_61h.csv}\\
    \input{Tables2/table_bline_e4_2_60h.csv}\\
    \input{Tables2/table_bline_e4_2_52h.csv}\\
    \input{Tables2/table_bline_e4_2_50h.csv}\\
    \input{Tables2/table_bline_e4_2_34h.csv}\\
    \input{Tables2/table_bline_e4_2_30h.csv}\\
    \bottomrule
\end{tabularx}
\begin{tablenotes}[flushleft]
\item \emph{Notes.} ***, **, * refers to statistical significance at the 1, 5, and 10\% level. Significance levels are computed using the bootstrap method. Hypothesis testing in this context is nonstandard, as it involves estimating proportions. We define significance at the $\alpha$ level as the case where more than $(1 - \alpha) \times 100\%$ of the bootstrap samples yield estimates greater than 0.0001.
\end{tablenotes}
\end{threeparttable}
\end{center}
\end{singlespace}
\end{table}

The estimates provide clearly the kind of information we need about how many track shifts occurred and which ones. This shows in our view the added value of the approach we propose to analyze this data effectively, on top of the previously presented threshold effects. The first row refers to the ``$+ 2$'' threshold for students with an initial \emph{havo} recommendation. We estimate that for these students, between 6.1 and 7.6\% was enrolled on a higher track by virtue of having a test score just above the threshold. Bounds become more important when we want to break this down into the different kinds of track shifts that take place. Based on our estimates, for example, we cannot be sure that the track shift $L_1\rightarrow H_1$ has actually occurred. For students with an initial \emph{havo/vwo} recommendation, we essentially reach point identification: 11\% of students experienced a shift from first year enrollment in a mixed \emph{havo/vwo} track to a single track \emph{vwo}.

These bounds on first year enrollment effects are the basis for interpreting the relevance of tracking effects for marginally assigned students. In column (1) and (2) of Table \ref{tab:table_bline_e4_1} we present upper and lower bound estimates of the fraction of students for whom being reassigned to a higher track, yields positive enrollment effects after four years of secondary education. We refer to these types of students as \emph{Trapped in Track}, as they are, provided they are upgraded to a higher track, trapped in a track that is too low for their capacities. In columns (3) and (4) of Table \ref{tab:table_bline_e4_1} we present upper and lower bound estimates on the fraction of students for whom being reassigned to a higher track, yields negative enrollment effects after four years of secondary school. We refer to these types of students as \emph{Slow Starters}, as they explicitly benefit from starting on a lower track.

In addition to the fraction of \emph{Trapped in Track} and \emph{Slow Starters}, we also present the Net effect, which is the difference between the fraction of \emph{Trapped in Track} and \emph{Slow Starters}. And, a somewhat new concept, the Total effect, which is the sum of the fraction of \emph{Trapped in Track} and \emph{Slow Starters}. The Total effect measures for how many students the outcome is effected, either positively or negatively, by the change in enrollment in the first year of secondary education. 

The results clearly indicate the importance of students experiencing positive effects of starting secondary education on a higher track. In a number of important cases, the fraction of \emph{Trapped in Track} students has a lower bound significantly larger than zero. Bounds on them are also often reasonably tight. The lower bounds on the \emph{Slow Starters} tend to be zero or close to it. They are also all statistically insignificant. 
At the same time, we cannot exclude the existence of \emph{Slow Starters} either. 

Another way of interpreting these results is that in general, the lower bound on the \emph{Trapped in Track} is close to (or equal to) the lower bound on the Total effect (the sum of the \emph{Trapped in Track} and \emph{Slow Starters}). This indicates that when \emph{Trapped in Track} is at its minimum, there cannot be any \emph{Slow Starters}. In other words, if there are \emph{Slow Starters}, there must be also be more additional \emph{Trapped in Track}.

We can now combine the results from Table \ref{tab:first_stage} and \ref{tab:table_bline_e4_1} to conclude that often a large share of the marginally assigned students are affected by the reassignment. For example, for students with an initial \emph{havo/vwo} recommendation, we can derive that between 37\% ($=100\%\times 4.1/11$) and 100\% ($=100\%\times 11/11$) of students are affected by the reassignment. For students at the $+ 2$ threshold with an initial \emph{havo} recommendation we can conclude that at least 50\% of students is affected. In fact, even more specifically, at least 50\% of these marginally assigned students are \emph{Trapped in Track}.\footnote{These lower bounds on these local average effects can be computed by dividing the lower bounds (from the \emph{Trapped in Track}, the \emph{Slow Starters} or the \emph{Total Effect}) presented in Table \ref{tab:table_bline_e4_1} by the upper bounds of the first stage, presented in Table \ref{tab:first_stage}. These quantities can have tighter bounds in principle as we could potentially compute associated first stage parameters for each Total effect, for example.} In  Appendix \ref{app:het_income} we show that these results appear more strongly and regularly in the low income subpopulation of our data.

\begin{table}[hbt!]
\begin{singlespace}
    \begin{center}
    \begin{threeparttable}
    \caption{\label{tab:table_bline_e4_1} Estimated fractions of students with positive (\emph{Trapped in Track}) and negative (\emph{Slow Starter}) effects of a positive change in first-year secondary school track enrollment, on \textbf{secondary school track enrollment four years after the start of secondary education}. Columns 5–6 report bounds on the difference between these fractions, while columns 7–8 report bounds on their sum. The results apply to the primary school graduation cohorts of 2014/15 to 2018/19.}
	\begin{tabularx}{1\textwidth}{@{\extracolsep{\fill}} lcccccccc}
		\toprule
         &\multicolumn{2}{c}{\emph{Trapped in Track}}&\multicolumn{2}{c}{\emph{Slow Starter}}&\multicolumn{2}{c}{Net EFFECT}&\multicolumn{2}{c}{Total EFFECT} \\ 
         \cmidrule(lr){2-5}\cmidrule(lr){6-7}\cmidrule(lr){8-9}
        &LB&UB&LB&UB&LB&UB&LB&UB\\    
            \cmidrule(lr){2-9}
            &(1)&(2)&(3)&(4)&(5)&(6)&(7)&(8)\\
		\cmidrule(lr){2-9}
    
    \input{Tables2/table_bline_e4_1_60w.csv}\\
    \input{Tables2/table_bline_e4_1_52w.csv}\\
    \input{Tables2/table_bline_e4_1_50w.csv}\\
    \input{Tables2/table_bline_e4_1_34w.csv}\\
    \input{Tables2/table_bline_e4_1_30w.csv}\\
    \input{Tables2/table_bline_e4_1_61h.csv}\\
    \input{Tables2/table_bline_e4_1_60h.csv}\\
    \input{Tables2/table_bline_e4_1_52h.csv}\\
    \input{Tables2/table_bline_e4_1_50h.csv}\\
    \input{Tables2/table_bline_e4_1_34h.csv}\\
    \input{Tables2/table_bline_e4_1_30h.csv}\\
    \bottomrule
	\end{tabularx}
\begin{tablenotes}[flushleft]
\item \emph{Notes.} ***, **, * refers to statistical significance at the 1, 5, and 10\% level. Significance levels are computed using the bootstrap method. Hypothesis testing in this context is nonstandard, as it involves estimating proportions. We define significance at the $\alpha$ level as the case where more than $(1 - \alpha) \times 100\%$ of the bootstrap samples yield estimates greater than 0.0001.
\end{tablenotes}
\end{threeparttable}
\end{center}
\end{singlespace}
\end{table}

In the third or fourth year of secondary school, depending on the track level, students must choose majors (Dutch: \emph{profiel}). In a broad sense there are three categories of profiles arguably somewhat increasing in the extent to which they rely on STEM subjects. In a similar way, they provide easier, more direct access to certain, more restrictive tertiary education programs. It might be that students who are, by virtue of an exogenous upward shift in enrollment, are approaching high school graduation at a higher level, but are in fact also choosing (or being forced to choose) less competitive majors. In Appendix \ref{app:major} we present simple threshold effect results on these different majors, similar to the results we have presented in Section \ref{sec:threshold_effects}. We find that, generally, the effects on major choice is often not significantly affected. 

With an eye to the existing literature on the (un)importance of tracking it seems relevant to extend our Table \ref{tab:table_bline_e4_1} to investigate \emph{Trapped in Track} and \emph{Slow Starters} by type of track shift. One notable result from this table is that we tend to find \emph{Trapped in Track} from shifts from mixed tracks to single tracks, for example, from the mixed \emph{havo/vwo} track to the single \emph{vwo} track in the ``havo $+ 2$'' and ``havo/vwo $+ 1$'' rows. We also find some less conclusive evidence for \emph{Trapped in Track} who shift into a mixed \emph{havo/vwo} track, for example from \emph{havo}. These results call into question the policy debate in the Netherlands, which almost uncritically argues for more mixed-ability tracks and a postponement of the tracking decision. Our findings indicate that such changes are likely to have negative consequences for at least some students. 

\begin{landscape}
\begin{table}[!hbt]
\begin{singlespace}
    \begin{center}
    \begin{threeparttable}
    \caption{\label{tab:table_bline_e4_5} Estimated fractions of students with positive effects (\emph{Trapped in Track}) and negative effects (\emph{Slow Starter}) of a positive change in first year secondary school track enrollment on \textbf{secondary school track enrollment four years after the start of secondary education} by shift in first year track enrollment.}
	\begin{tabularx}{1.3\textwidth}{@{\extracolsep{\fill}} lcccccccccccc}
		\toprule
         &\multicolumn{6}{c}{\emph{Trapped in Track}}&\multicolumn{6}{c}{\emph{Slow Starters}}\\
         \cmidrule(lr){2-7}\cmidrule(lr){8-13}
         &\multicolumn{2}{c}{$L_1\rightarrow M_1$}&\multicolumn{2}{c}{$L_1\rightarrow H_1$}&\multicolumn{2}{c}{$M_1\rightarrow H_1$}
         &\multicolumn{2}{c}{$L_1\rightarrow M_1$}&\multicolumn{2}{c}{$L_1\rightarrow H_1$}&\multicolumn{2}{c}{$M_1\rightarrow H_1$}\\
         \cmidrule(lr){2-7}\cmidrule(lr){8-13}
        &LB&UB&LB&UB&LB&UB&LB&UB&LB&UB&LB&UB\\    
            \cmidrule(lr){2-13}
            &(1)&(2)&(3)&(4)&(5)&(6)&(7)&(8)&(9)&(10)&(11)&(12)\\
		\cmidrule(lr){2-13}
    \input{Tables2/table_bline_e4_8_60w.csv}\\
    \input{Tables2/table_bline_e4_8_52w.csv}\\
    \input{Tables2/table_bline_e4_8_50w.csv}\\
    \input{Tables2/table_bline_e4_8_34w.csv}\\
    \input{Tables2/table_bline_e4_8_30w.csv}\\
    \input{Tables2/table_bline_e4_8_61h.csv}\\
    \input{Tables2/table_bline_e4_8_60h.csv}\\
    \input{Tables2/table_bline_e4_8_52h.csv}\\
    \input{Tables2/table_bline_e4_8_50h.csv}\\
    \input{Tables2/table_bline_e4_8_34h.csv}\\
    \input{Tables2/table_bline_e4_8_30h.csv}\\
    \bottomrule
	\end{tabularx}
\begin{tablenotes}[flushleft]
\item \emph{Notes.} ***, **, * refers to statistical significance at the 1, 5, and 10\% level. Significance levels are computed using the bootstrap method. Hypothesis testing in this context is nonstandard, as it involves estimating proportions. We define significance at the $\alpha$ level as the case where more than $(1 - \alpha) \times 100\%$ of the bootstrap samples yield estimates greater than 0.0001.
\end{tablenotes}
\end{threeparttable}
\end{center}
\end{singlespace}
\end{table}
\end{landscape}

\subsection{The quality of track assignment}

The results presented so far has almost uniformly indicated positive tracking effects. Often, we find that among marginal students, at least 40\% benefit from the reassignment. That is, for students at the margin of being assigned to different track levels, the higher track tends to yield better outcomes. And these outcomes also tend to persist into the future (see Appendix \ref{app:long_term_effects} for these long term effects). Two other results so far are that track upgrading seems to occur across all thresholds and that not all the upgrading teachers that teachers do, are followed by parents and students. 

In section \ref{sec:model} we have presented a theoretical model to derive predictions based on a rational model of track assignment, under a variety of (pre)conditions. The main result of this exercise was that we might expect positive effects from upgrading at the $+ 1$ threshold as the incentives to assign conservatively have been removed and because the requirement to reassess the initial recommendation might lead to updated beliefs, based on new information (potentially derived from a high test score on the end-of-primary education test). In addition, we predicted that zero or slightly negative effects would drive the results at the $+ 2$ thresholds. This would be because both on the left and the right side of the $+ 2$ threshold, the main arguments for upgrading are psychological costs of not doing so (which would lead to a higher recommendation to the right of the threshold than would be preferred when these costs would not be there) and the nudge, which should only influence the decision when teachers are reasonably indifferent between two track levels. Reasonable indifference is plausible as it is considerably difficult to make such decisions with any level of certainty. 

From this perspective, the positive effects we estimate at the $+ 2$ thresholds in particular appear inconsistent with the rational model presented in \ref{sec:model}. That is, if teachers would in fact upgrade at the $+ 2$ threshold when they are indifferent between two track types, and/or whether they only upgrade because they are not confident enough to justify the decision not to upgrade to parents, our findings seem to indicate that teachers tend to assign conservatively at the margin. 

This conclusion however is still premature because teachers do not unilaterally decide on track enrollment. Parents and students have to form their own opinions about which school to attend and which track. There are two ways in which these processes could justify these positive effects at the $+ 2$ thresholds, while still maintaining the core of the model predictions. One is, that there are ways beyond the recommendation of enrolling at a track level above the recommended level. This route is considerably difficult for parents and students, as secondary schools usually do not allow this. But it is conceivable to us that, at times, secondary schools might upgrade students themselves as classes need to be filled. In such cases it is also conceivable that they would start doing so with students with high test scores. Another rationale for positive effects (while zero, or negative effects were anticipated) is that parents filter out the \emph{Slow Starters} and prevent them from enrolling at the recommended level that they deem too high.  

In Table \ref{tab:shift_shift} we use our empirical framework to arrange four groups of students, based on whether the student's recommendation was upgraded and whether they started secondary education on a higher track. The columns (1) and (2) refer to the group that receive an upgrade in the recommendation and also started secondary education on a higher level. The columns (5) and (6) refer to students who did not receive an upgraded recommendation, but still were upgraded to a higher track level by virtue of a test score right above the threshold. Somewhat surprisingly we cannot exclude the possibility that all the shifting in first year enrollment, took place without having received an upgrade in the recommendation. This result however does support our decision to allow for the possibility that there are effects on first year enrollment without the upgrade. Restricting this would be inconsistent with the data and potentially produce unreliable results. 

\begin{table}[!h]
\begin{singlespace}
    \begin{center}
    \begin{threeparttable}
    \caption{\label{tab:shift_shift} Shifts in recommended track level combined with shifts in first year track enrollment}
	\begin{tabularx}{1\textwidth}{@{\extracolsep{\fill}} lcccccccc}
		\toprule
         &\multicolumn{2}{c}{shift - shift}&\multicolumn{2}{c}{shift - no shift}&\multicolumn{2}{c}{no shift - shift}&\multicolumn{2}{c}{no shift - no shift}\\ 
         \cmidrule(lr){2-5}\cmidrule(lr){6-9}
        &LB&UB&LB&UB&LB&UB&LB&UB\\    
            \cmidrule(lr){2-9}
            &(1)&(2)&(3)&(4)&(5)&(6)&(7)&(8)\\
		\cmidrule(lr){2-9}
    
    \input{Tables2/table_bline_e4_3_61h.csv}\\
    \input{Tables2/table_bline_e4_3_60w.csv}\\
    \input{Tables2/table_bline_e4_3_60h.csv}\\
    \input{Tables2/table_bline_e4_3_52w.csv}\\
    \input{Tables2/table_bline_e4_3_52h.csv}\\
    \input{Tables2/table_bline_e4_3_50w.csv}\\
    \input{Tables2/table_bline_e4_3_50h.csv}\\
    \input{Tables2/table_bline_e4_3_34w.csv}\\
    \input{Tables2/table_bline_e4_3_34h.csv}\\
    \input{Tables2/table_bline_e4_3_30w.csv}\\
    \input{Tables2/table_bline_e4_3_30h.csv}\\
    \bottomrule
	\end{tabularx}
\begin{tablenotes}[flushleft]
\item \emph{Notes.} ***, **, * refers to statistical significance at the 1, 5, and 10\% level. Significance levels are computed using the bootstrap method. Hypothesis testing in this context is nonstandard, as it involves estimating proportions. We define significance at the $\alpha$ level as the case where more than $(1 - \alpha) \times 100\%$ of the bootstrap samples yield estimates greater than 0.0001.
\end{tablenotes}
\end{threeparttable}
\end{center}
\end{singlespace}
\end{table}

The results in Table \ref{tab:shift_shift} thus show that we cannot cleanly trace the tracking effects in Section \ref{sec:tracking_effects} to the decisions that were made by teachers. Within the context of our specification we cannot be sure whether the \emph{Trapped in Track} students, who benefited from being reassigned, were actually upgraded by their teachers. In Table \ref{tab:table_bline_e4_9} this finding is confirmed. In Table \ref{tab:table_bline_e4_9} we estimate bounds on the fractions of \emph{Trapped in Track}, \emph{Slow Starters}, \emph{Always Low}, \emph{Always Middle} and \emph{Always High} for only those students that receive an upgrade in the recommendation and also started secondary education on a higher level.

\begin{landscape}
\begin{table}[!hbt]
\begin{singlespace}
    \begin{center}
    \begin{threeparttable}
    \caption{\label{tab:table_bline_e4_9} Bounds on the fraction of student types, that can be linked to the shift in the teacher's recommendation}
	\begin{tabularx}{1.3\textwidth}{@{\extracolsep{\fill}} lcccccccccc}
		\toprule
         &\multicolumn{2}{c}{shift-shift \emph{TT}}&\multicolumn{2}{c}{shift-shift \emph{SS}}&\multicolumn{2}{c}{shift-shift \emph{AL}}&\multicolumn{2}{c}{shift-shift \emph{AM}}
         &\multicolumn{2}{c}{shift-shift \emph{AH}}\\
         \cmidrule(lr){2-11}
        &LB&UB&LB&UB&LB&UB&LB&UB&LB&UB\\    
            \cmidrule(lr){2-11}
            &(1)&(2)&(3)&(4)&(5)&(6)&(7)&(8)&(9)&(10)\\
		\cmidrule(lr){2-11}
    \input{Tables2/table_bline_e4_9_61h.csv}\\
    \input{Tables2/table_bline_e4_9_60w.csv}\\
    \input{Tables2/table_bline_e4_9_60h.csv}\\
    \input{Tables2/table_bline_e4_9_52w.csv}\\
    \input{Tables2/table_bline_e4_9_52h.csv}\\
    \input{Tables2/table_bline_e4_9_50w.csv}\\
    \input{Tables2/table_bline_e4_9_50h.csv}\\
    \input{Tables2/table_bline_e4_9_34w.csv}\\
    \input{Tables2/table_bline_e4_9_34h.csv}\\
    \input{Tables2/table_bline_e4_9_30w.csv}\\
    \input{Tables2/table_bline_e4_9_30h.csv}\\
    \bottomrule
	\end{tabularx}
\begin{tablenotes}[flushleft]
\item \emph{Notes.} ***, **, * refers to statistical significance at the 1, 5, and 10\% level. Significance levels are computed using the bootstrap method. Hypothesis testing in this context is nonstandard, as it involves estimating proportions. We define significance at the $\alpha$ level as the case where more than $(1 - \alpha) \times 100\%$ of the bootstrap samples yield estimates greater than 0.0001.
\end{tablenotes}
\end{threeparttable}
\end{center}
\end{singlespace}
\end{table}
\end{landscape}

It is however still possible to draw conclusions that are grounded in our empirical results. First, at the margin of the assignment process, at least 40\% of students benefit from being reassigned. This shows clearly that at least at the margins of assignment, the assignment process is ``difficult'' in the sene that assignment matters. The results show that at the margin the process is at least noisy, with a lot of students who are ex post wrongly assigned. 

Also, the model predicts that at the $+ 2$ thresholds, the effects of the upgrades should have zero, or somewhat negative effects in expectation. This means that those who are reassigned should be ex ante more likely to be \emph{Slow Starter} than \emph{Trapped in Track}. We do not see this in the data. Instead we are able to detect many more \emph{Trapped in Track}. The teacher's assignment can still however be in line with the model, and hence, be outcome maximizing, when parents filter out the \emph{Slow Starter}, so that only the \emph{Trapped in Track} actually start secondary education on a higher track level. This scenario is one in which the teachers are following the model and where the parents have superior knowledge about the ability level of these students. Also, when these \emph{Slow Starters} are there, unobserved by us, the process of assignment is even more noisy that we can now establish.

We cannot exclude the possibility that the \emph{Trapped in Track} we find in section \ref{sec:tracking_effects} were not even upgraded by their teachers. In that case, teachers seem to be ex-ante quite confident for these students that starting on a higher track level would not be a good idea. At first glance, therefore, it appears surprising that those who managed to start on a higher track level without the upgraded recommendation, are very likely to benefit from it. If this process would be somewhat random, for example, due to administrative reasons (such as class-size restrictions), there might also be a lot of \emph{Trapped in Track} who did not have the opportunity to start on a higher level in order to benefit from it. 

%% file: Tables2/table_bline_e4_2_60w.csv
havo + 2&0.061***&0.076***&0.010&0.025***&0.000&0.015**&0.036***&0.051***

%% file: Tables2/table_bline_e4_2_52w.csv
vmbo-gt/havo + 2&0.094***&0.101***&0.000&0.007***&0.000&0.007**&0.087***&0.094***

%% file: Tables2/table_bline_e4_2_50w.csv
vmbo-gt + 2&0.060***&0.089***&0.015***&0.044***&0.000&0.029***&0.016**&0.045***

%% file: Tables2/table_bline_e4_2_34w.csv
vmbo-kb/gt + 2&0.118***&0.168***&0.006&0.056***&0.000&0.050***&0.062***&0.112***

%% file: Tables2/table_bline_e4_2_30w.csv
vmbo-kb + 2&0.093***&0.165***&0.012**&0.085***&0.000&0.073***&0.008*&0.081***

%% file: Tables2/table_bline_e4_2_61h.csv
havo/vwo + 1&0.111***&0.112***&0.000&0.001&0.000&0.001&0.110***&0.111***

%% file: Tables2/table_bline_e4_2_60h.csv
havo + 1&0.027***&0.030***&0.023***&0.026***&0.000&0.002&0.002&0.004***

%% file: Tables2/table_bline_e4_2_52h.csv
vmbo-gt/havo + 1&0.077***&0.105***&0.004*&0.032***&0.000&0.028**&0.046***&0.074***

%% file: Tables2/table_bline_e4_2_50h.csv
vmbo-gt + 1&0.015***&0.015***&0.015***&0.015***&0.000&0.000&0.001&0.001*

%% file: Tables2/table_bline_e4_2_34h.csv
vmbo-kb/gt + 1&0.066***&0.097***&0.010&0.040***&0.000&0.030*&0.026**&0.056***

%% file: Tables2/table_bline_e4_2_30h.csv
vmbo-kb + 1&0.048***&0.062***&0.006&0.020**&0.000&0.014&0.028**&0.042***

%% file: Tables2/table_bline_e4_1_60w.csv
havo + 2&0.042***&0.061***&0.000&0.018***&0.024**&0.051***&0.042***&0.076***

%% file: Tables2/table_bline_e4_1_52w.csv
vmbo-gt/havo + 2&0.004&0.041***&0.010&0.048***&-0.014&0.001&0.014***&0.089***

%% file: Tables2/table_bline_e4_1_50w.csv
vmbo-gt + 2&0.028***&0.058***&0.003&0.032***&0.001&0.034***&0.031***&0.089***

%% file: Tables2/table_bline_e4_1_34w.csv
vmbo-kb/gt + 2&0.037*&0.074***&0.010&0.045***&0.020&0.043*&0.047***&0.119***

%% file: Tables2/table_bline_e4_1_30w.csv
vmbo-kb + 2&0.078***&0.096***&0.000&0.040***&0.051***&0.089***&0.078***&0.137***

%% file: Tables2/table_bline_e4_1_61h.csv
havo/vwo + 1&0.041***&0.083***&0.000&0.030***&0.033***&0.061***&0.041***&0.112***

%% file: Tables2/table_bline_e4_1_60h.csv
havo + 1&0.010**&0.022***&0.000&0.002***&0.010*&0.020**&0.010***&0.024***

%% file: Tables2/table_bline_e4_1_52h.csv
vmbo-gt/havo + 1&0.048***&0.079***&0.000&0.028***&0.035***&0.062***&0.048***&0.105***

%% file: Tables2/table_bline_e4_1_50h.csv
vmbo-gt + 1&0.006&0.010***&0.005&0.006***&0.001&0.005&0.011***&0.015***

%% file: Tables2/table_bline_e4_1_34h.csv
vmbo-kb/gt + 1&0.001&0.054***&0.000&0.033***&-0.001&0.024&0.001***&0.087***

%% file: Tables2/table_bline_e4_1_30h.csv
vmbo-kb + 1&0.013&0.028***&0.004&0.028***&-0.005&0.010&0.017***&0.056***

%% file: Tables2/table_bline_e4_8_60w.csv
havo + 2&0.010&0.025***&0.000&0.015**&0.017**&0.048***&0.000&0.015**&0.000&0.003**&0.000&0.018***

%% file: Tables2/table_bline_e4_8_52w.csv
vmbo-gt/havo + 2&0.000&0.003**&0.000&0.003**&0.001&0.038***&0.000&0.003&0.000&0.003&0.007&0.045***

%% file: Tables2/table_bline_e4_8_50w.csv
vmbo-gt + 2&0.004&0.035***&0.000&0.024***&0.000&0.044***&0.003&0.031***&0.000&0.010***&0.000&0.021***

%% file: Tables2/table_bline_e4_8_34w.csv
vmbo-kb/gt + 2&0.000&0.048***&0.000&0.048***&0.000&0.072***&0.004&0.018***&0.000&0.016**&0.000&0.027***

%% file: Tables2/table_bline_e4_8_30w.csv
vmbo-kb + 2&0.006&0.079***&0.000&0.073***&0.001&0.079***&0.000&0.031***&0.000&0.008***&0.000&0.009***

%% file: Tables2/table_bline_e4_8_61h.csv
havo/vwo + 1&0.000&0.001&0.000&0.001&0.040***&0.083***&0.000&0.001&0.000&0.001&0.000&0.030***

%% file: Tables2/table_bline_e4_8_60h.csv
havo + 1&0.009**&0.020***&0.000&0.002&0.001&0.003**&0.000&0.002*&0.000&0.001&0.000&0.002*

%% file: Tables2/table_bline_e4_8_52h.csv
vmbo-gt/havo + 1&0.000&0.014**&0.000&0.014*&0.034***&0.074***&0.000&0.028**&0.000&0.012*&0.000&0.012*

%% file: Tables2/table_bline_e4_8_50h.csv
vmbo-gt + 1&0.006&0.010**&0.000&0.000&0.000&0.000&0.005&0.005**&0.000&0.000&0.000&0.000

%% file: Tables2/table_bline_e4_8_34h.csv
vmbo-kb/gt + 1&0.000&0.000&0.000&0.000&0.001&0.054***&0.000&0.030**&0.000&0.025*&0.000&0.025**

%% file: Tables2/table_bline_e4_8_30h.csv
vmbo-kb + 1&0.000&0.014&0.000&0.014&0.000&0.024**&0.000&0.000&0.000&0.000&0.004&0.028***

%% file: Tables2/table_bline_e4_3_61h.csv
havo/vwo + 1&0.000&0.082***&0.075***&0.159***&0.030***&0.112***&0.676***&0.815***

%% file: Tables2/table_bline_e4_3_60w.csv
havo + 2&0.000&0.060***&0.022***&0.133***&0.016***&0.076***&0.742***&0.903***

%% file: Tables2/table_bline_e4_3_60h.csv
havo + 1&0.000&0.029***&0.041***&0.071***&0.001***&0.030***&0.900***&0.930***

%% file: Tables2/table_bline_e4_3_52w.csv
vmbo-gt/havo + 2&0.000&0.083***&0.015***&0.132***&0.018***&0.101***&0.518***&0.886***

%% file: Tables2/table_bline_e4_3_52h.csv
vmbo-gt/havo + 1&0.000&0.092***&0.031***&0.123***&0.014***&0.105***&0.709***&0.869***

%% file: Tables2/table_bline_e4_3_50w.csv
vmbo-gt + 2&0.000&0.076***&0.006***&0.117***&0.013***&0.089***&0.763***&0.913***

%% file: Tables2/table_bline_e4_3_50h.csv
vmbo-gt + 1&0.000&0.015***&0.046***&0.063***&0.001***&0.015***&0.917***&0.939***

%% file: Tables2/table_bline_e4_3_34w.csv
vmbo-kb/gt + 2&0.000&0.157***&0.034***&0.207***&0.011***&0.168***&0.313***&0.814***

%% file: Tables2/table_bline_e4_3_34h.csv
vmbo-kb/gt + 1&0.000&0.080***&0.048***&0.129***&0.003**&0.097***&0.760***&0.878***

%% file: Tables2/table_bline_e4_3_30w.csv
vmbo-kb + 2&0.000&0.102***&0.022***&0.129***&0.021***&0.164***&0.706***&0.856***

%% file: Tables2/table_bline_e4_3_30h.csv
vmbo-kb + 1&0.006*&0.054***&0.032***&0.081***&0.008***&0.056***&0.915***&0.966***

%% file: Tables2/table_bline_e4_9_61h.csv
havo/vwo + 1&0.000&0.074***&0.000&0.026***&0.000&0.004***&0.000&0.024***&0.000&0.023***

%% file: Tables2/table_bline_e4_9_60w.csv
havo + 2&0.000&0.053***&0.000&0.017***&0.000&0.002***&0.000&0.026***&0.000&0.015***

%% file: Tables2/table_bline_e4_9_60h.csv
havo + 1&0.000&0.022***&0.000&0.002**&0.000&0.003**&0.000&0.019***&0.000&0.001*

%% file: Tables2/table_bline_e4_9_52w.csv
vmbo-gt/havo + 2&0.000&0.041***&0.000&0.033***&0.000&0.005***&0.000&0.027***&0.000&0.050***

%% file: Tables2/table_bline_e4_9_52h.csv
vmbo-gt/havo + 1&0.000&0.075***&0.000&0.028***&0.000&0.001**&0.000&0.025***&0.000&0.045***

%% file: Tables2/table_bline_e4_9_50w.csv
vmbo-gt + 2&0.000&0.049***&0.000&0.027***&0.000&0.000**&0.000&0.042***&0.000&0.028***

%% file: Tables2/table_bline_e4_9_50h.csv
vmbo-gt + 1&0.000&0.010***&0.000&0.005***&0.000&0.003**&0.000&0.000**&0.000&0.000**

%% file: Tables2/table_bline_e4_9_34w.csv
vmbo-kb/gt + 2&0.000&0.074***&0.000&0.030***&0.000&0.002*&0.000&0.036***&0.000&0.093***

%% file: Tables2/table_bline_e4_9_34h.csv
vmbo-kb/gt + 1&0.000&0.050***&0.000&0.031***&0.000&0.003&0.000&0.019***&0.000&0.047***

%% file: Tables2/table_bline_e4_9_30w.csv
vmbo-kb + 2&0.000&0.084***&0.000&0.018***&0.000&0.005***&0.000&0.026***&0.000&0.056***

%% file: Tables2/table_bline_e4_9_30h.csv
vmbo-kb + 1&0.000&0.028***&0.000&0.021***&0.000&0.016**&0.000&0.010***&0.000&0.014***

%% file: Conclusion.tex
\section{Conclusion}\label{sec:conclusion}

In this paper we study the quality of track assignment for students at the margin of being assigned to different tracks. The concept of assignment quality has received little attention in the context of educational tracking, whereas biases in the assignment process -- assignment to a track that is not maximizing outcomes in expectation -- could help explain some of the mixed results found in the literature. In particular, using a model of optimal track assignment under uncertainty, we predict that optimal (outcome maximizing) assignment implies, in some of our settings, weakly negative of zero average tracking effects for marginally assigned students.

To test this prediction, we study tracking effects for students who are at the margin of being assigned to different tracks. In the Netherlands, track assignments are based on a decision process in which teachers first, and parents second, determine the starting track level at which students start secondary education around age 12. For primary students in 6th grade, teachers determine an initial track recommendation, after which students take a standardized school-leavers test. When students score above certain threshold levels on this test, the teacher has to consider an upward revision of the track recommendation. 

To analyze tracking effects we develop a flexible causal approach, which for our purpose is embedded within the context of a regression discontinuity design. The approach allows for the separation, organization, and partial identification of the various different tracking effects underlying the overall estimated effects at the test score cutoffs. Our results indicate substantial tracking effects: between 40\% and 100\% of marginally assigned students are positive or negatively affected by enrolling in a higher track. Most tracking effects are positive, however, with students benefiting from being placed in a higher, more demanding track. While based on the current analysis we cannot reject the hypothesis that teacher assignments are unbiased, this result seems only consistent with a significant degree of noise. We discuss that parental decisions, whether to follow or deviate from teacher recommendations, may help reducing this noise.

%% file: Appendix.tex
\section{List of cutoff levels}\label{app:Cutoffs}

\begin{table}[H]
	\begin{center}
		\begin{threeparttable}
	    \caption{\label{tab:toetsadvies} The test-based recommendation is a mapping from the achievement test score to a track level}
			\begin{tabularx}{1\textwidth}{@{\extracolsep{\fill}} lccccc}
			\toprule
			& (1)& (2)& (3) & (4)& (5) \\ 
    	\cmidrule(lr){2-6}
     &\multicolumn{1}{c}{2014/15}&\multicolumn{1}{c}{2015/16}&\multicolumn{1}{c}{2016/17}&\multicolumn{1}{c}{2017/18}&\multicolumn{1}{c}{2018/19}\\
     &[min-max]&[min-max]&[min-max]&[min-max]&[min-max]\\
    \cmidrule(lr){2-6}
    \input{Tables/toetsadvies_fixed}\\
    \bottomrule
    \end{tabularx}
    \begin{tablenotes}[flushleft]
	\item \emph{Notes.} Table reports the range of test scores (from minimum to maximum) that map into a track level, which we refer to as the test-based track recommendation. The minimum scores are the test score cutoffs. These numbers reflect the test scores used by the \emph{Cito} end-of-primary education achievement test. The highest and the lowest possible scores are 550 and 501 respectively. The average score in the population of primary school 6th graders is about 535. 
\end{tablenotes}
\end{threeparttable}
\end{center}
\end{table}

\section{Coding rules for the mapping of track types to placeholders $L$, $M$ and $H$}\label{app:coding_rules}

\begin{table}[H]
	\begin{center}
		\begin{threeparttable}
	    \caption{\label{tab:coding_rules} Coding rules}
			\begin{tabularx}{1\textwidth}{@{\extracolsep{\fill}} lccc}
			\toprule
    \input{Tables2/table_rules0.csv}\\\\
    \input{Tables2/table_rules1.csv}\\\\
    \input{Tables2/table_rules4.csv}\\
    \bottomrule
    \end{tabularx}
\end{threeparttable}
\end{center}
\end{table}

%% file: Tables/toetsadvies_fixed.tex
 &  &  &  &  &   \\
vmbo-bl/vmbo-kl 	&   			& [519 - 525] & [519 - 525] & [519 - 525] & \\
vmbo-kl 		& [524 - 528] 		& [526 - 528] & [526 - 528] & [526 - 528] & \\
vmbo-kl/vmbo-gt		&			&	      &	 	    &		  & [525 - 533]\\	
vmbo-gt 		& [529 - 536] 		& [529 - 532] & [529 - 532] & [529 - 532] & \\
vmbo-gt/havo 		&   			& [533 - 536] & [533 - 536] & [533 - 536] & [533 - 539]\\
havo 			& [537 - 544] 		& [537 - 539] & [537 - 539] & [537 - 539] & \\
havo/vwo 		&  			& [540 - 544] & [540 - 544] & [540 - 544] & [540 - 544]\\
vwo 			& [545 - 550] 		& [545 - 550] & [545 - 550] & [545 - 550] & [545 - 550]\\

%% file: Tables2/table_rules0.csv
&$L_0$&$M_0$&$H_0$\\\cmidrule{2-4}
havo/vwo&$\leq$havo/vwo&vwo&\\
havo&$\leq$havo&havo/vwo&$\geq$vwo\\
vmbo-gt/havo&$\leq$vmbo-gt/havo&havo&$\geq$havo/vwo\\
vmbo-gt&$\leq$vmbo-gt&vmbo-gt/havo&$\geq$havo\\
vmbo-kb/gt&$\leq$vmbo-kb/gt&vmbo-gt&$\geq$vmbo-gt/havo\\
vmbo-kb&$\leq$vmbo-kb&vmbo-kb/gt&$\geq$vmbo-gt\\
vmbo-bb/kb&$\leq$vmbo-bb/kb&vmbo-kb&$\geq$vmbo-kb/gt\\
vmbo-bb&$\leq$vmbo-bb&vmbo-bb/kb&$\geq$vmbo-bb/kb

%% file: Tables2/table_rules1.csv
&$L_1$&$M_1$&$H_1$\\\cmidrule{2-4}
havo/vwo&$\leq$havo&havo/vwo&$\geq$vwo\\
havo&$\leq$havo&havo/vwo&$\geq$vwo\\
vmbo-gt/havo&$\leq$vmbo-gt&vmbo-gt/havo&$\geq$havo\\
vmbo-gt&$\leq$vmbo-gt&vmbo-gt/havo&$\geq$havo\\
vmbo-kb/gt&$\leq$vmbo-kb&vmbo-kb/gt&$\geq$vmbo-gt\\
vmbo-kb&$\leq$vmbo-kb&vmbo-kb/gt&$\geq$vmbo-gt\\
vmbo-bb/kb&$\leq$vmbo-bb&vmbo-bb/kb&$\geq$vmbo-kb\\
vmbo-bb&$\leq$vmbo-bb&vmbo-bb/kb&$\geq$vmbo-kb

%% file: Tables2/table_rules4.csv
&$L_4$&$M_4$&$H_4$\\\cmidrule{2-4}
havo/vwo&$\leq$vmbo-gt&havo&$\geq$vwo\\
havo&$\leq$vmbo-gt&havo&$\geq$vwo\\
vmbo-gt/havo&$\leq$vmbo-kb&vmbo-gt&$\geq$havo\\
vmbo-gt&$\leq$vmbo-kb&vmbo-gt&$\geq$havo\\
vmbo-kb/gt&$\leq$vmbo-bb&vmbo-kb&$\geq$vmbo-gt\\
vmbo-kb&$\leq$vmbo-bb&vmbo-kb&$\geq$vmbo-gt\\
vmbo-bb/kb&$<$vmbo-bb&vmbo-bb&$\geq$vmbo-kb\\
vmbo-bb&$<$vmbo-bb&vmbo-bb&$\geq$vmbo-kb

%% file: Appendix_results.tex
\section{Heterogenous effects by parental income levels}\label{app:het_income}

We have estimated the parameters of Table \ref{tab:table_bline_e4_1} separately for low income (below median income, conditional on the initial recommendation) and high income (above median income, conditional on the initial recommendation). The results are presented in Tables \ref{tab:table_arm_e4_1} and \ref{tab:table_rijk_e4_1} respectively. For this draft, the results do not have the markers for statistical significance yet. We can see however that the results appear more strongly and regularly for the low income subpopulation. 

\begin{table}[hbt!]
\begin{singlespace}
    \begin{center}
    \begin{threeparttable}
    \caption{\label{tab:table_arm_e4_1} Estimated fractions of students with positive (\emph{Trapped in Track}) and negative (\emph{Slow Starter}) effects of a positive change in first-year secondary school track enrollment, on \textbf{secondary school track enrollment four years after the start of secondary education}. Results refer to students with \textbf{below median parental income}. Columns 5–6 report bounds on the difference between these fractions, while columns 7–8 report bounds on their sum. The results apply to the primary school graduation cohorts of 2014/15 to 2018/19.}
	\begin{tabularx}{1\textwidth}{@{\extracolsep{\fill}} lcccccccc}
		\toprule
         &\multicolumn{2}{c}{\emph{Trapped in Track}}&\multicolumn{2}{c}{\emph{Slow Starter}}&\multicolumn{2}{c}{Net EFFECT}&\multicolumn{2}{c}{Total EFFECT} \\ 
         \cmidrule(lr){2-5}\cmidrule(lr){6-7}\cmidrule(lr){8-9}
        &LB&UB&LB&UB&LB&UB&LB&UB\\    
            \cmidrule(lr){2-9}
            &(1)&(2)&(3)&(4)&(5)&(6)&(7)&(8)\\
		\cmidrule(lr){2-9}
    
    \input{Tables2/table_arm_e4_1_60w.csv}\\
    \input{Tables2/table_arm_e4_1_52w.csv}\\
    \input{Tables2/table_arm_e4_1_50w.csv}\\
    \input{Tables2/table_arm_e4_1_34w.csv}\\
    \input{Tables2/table_arm_e4_1_30w.csv}\\
    \input{Tables2/table_arm_e4_1_61h.csv}\\
    \input{Tables2/table_arm_e4_1_60h.csv}\\
    \input{Tables2/table_arm_e4_1_52h.csv}\\
    \input{Tables2/table_arm_e4_1_50h.csv}\\
    \input{Tables2/table_arm_e4_1_34h.csv}\\
    \input{Tables2/table_arm_e4_1_30h.csv}\\
    \bottomrule
	\end{tabularx}
\begin{tablenotes}[flushleft]
\item \emph{Notes.} +, indicates that markers for statistical significance are not yet obtained.
\end{tablenotes}
\end{threeparttable}
\end{center}
\end{singlespace}
\end{table}

\begin{table}[hbt!]
\begin{singlespace}
    \begin{center}
    \begin{threeparttable}
    \caption{\label{tab:table_rijk_e4_1} Estimated fractions of students with positive (\emph{Trapped in Track}) and negative (\emph{Slow Starter}) effects of a positive change in first-year secondary school track enrollment, on \textbf{secondary school track enrollment four years after the start of secondary education}. Results refer to students with \textbf{above median parental income}. Columns 5–6 report bounds on the difference between these fractions, while columns 7–8 report bounds on their sum. The results apply to the primary school graduation cohorts of 2014/15 to 2018/19.}
	\begin{tabularx}{1\textwidth}{@{\extracolsep{\fill}} lcccccccc}
		\toprule
         &\multicolumn{2}{c}{\emph{Trapped in Track}}&\multicolumn{2}{c}{\emph{Slow Starter}}&\multicolumn{2}{c}{Net EFFECT}&\multicolumn{2}{c}{Total EFFECT} \\ 
         \cmidrule(lr){2-5}\cmidrule(lr){6-7}\cmidrule(lr){8-9}
        &LB&UB&LB&UB&LB&UB&LB&UB\\    
            \cmidrule(lr){2-9}
            &(1)&(2)&(3)&(4)&(5)&(6)&(7)&(8)\\
		\cmidrule(lr){2-9}
    
    \input{Tables2/table_rijk_e4_1_60w.csv}\\
    \input{Tables2/table_rijk_e4_1_52w.csv}\\
    \input{Tables2/table_rijk_e4_1_50w.csv}\\
    \input{Tables2/table_rijk_e4_1_34w.csv}\\
    \input{Tables2/table_rijk_e4_1_30w.csv}\\
    \input{Tables2/table_rijk_e4_1_61h.csv}\\
    \input{Tables2/table_rijk_e4_1_60h.csv}\\
    \input{Tables2/table_rijk_e4_1_52h.csv}\\
    \input{Tables2/table_rijk_e4_1_50h.csv}\\
    \input{Tables2/table_rijk_e4_1_34h.csv}\\
    \input{Tables2/table_rijk_e4_1_30h.csv}\\
    \bottomrule
	\end{tabularx}
\begin{tablenotes}[flushleft]
\item \emph{Notes.} +, indicates that markers for statistical significance are not yet obtained.
\end{tablenotes}
\end{threeparttable}
\end{center}
\end{singlespace}
\end{table}

\clearpage

\section{Effects on secondary school major choice}\label{app:major}

\begin{table}[!htb]
\begin{singlespace}
    \begin{center}
    \begin{threeparttable}
    \caption{\label{tab:thresholdeffects_major} Threshold effects secondary school major choice}
	\begin{tabularx}{1\textwidth}{@{\extracolsep{\fill}} lcccccc}
		\toprule
        &\multicolumn{2}{c}{\emph{Cultuur/Economie \&}} &\multicolumn{2}{c}{\emph{Natuur \& Gezondheid}}&\multicolumn{2}{c}{\emph{Natuur \& Techniek}}\\ 
         &\multicolumn{2}{c}{\emph{Maatschappij}} & & \\
&$\alpha_Y$&$\beta_Y$&$\alpha_Y$&$\beta_Y$&$\alpha_Y$&$\beta_Y$\\   
            \cmidrule(lr){2-7}
            &(1)&(2)&(3)&(4)&(5)&(6)\\
		\cmidrule(lr){2-7}
    \input{Tables2/table_rf_60w_baseline_major.csv}\\
    \input{Tables2/table_rf_52w_baseline_major.csv}\\
    \input{Tables2/table_rf_50w_baseline_major.csv}\\
    \input{Tables2/table_rf_34w_baseline_major.csv}\\
    \input{Tables2/table_rf_30w_baseline_major.csv}\\
    \input{Tables2/table_rf_22w_baseline_major.csv}\\
    \input{Tables2/table_rf_20w_baseline_major.csv}\\
    \input{Tables2/table_rf_61h_baseline_major.csv}\\
    \input{Tables2/table_rf_60h_baseline_major.csv}\\
    \input{Tables2/table_rf_52h_baseline_major.csv}\\
    \input{Tables2/table_rf_50h_baseline_major.csv}\\
    \input{Tables2/table_rf_34h_baseline_major.csv}\\
    \input{Tables2/table_rf_30h_baseline_major.csv}\\
    \input{Tables2/table_rf_22h_baseline_major.csv}\\
    \input{Tables2/table_rf_20h_baseline_major.csv}\\
    \bottomrule
	\end{tabularx}
\begin{tablenotes}[flushleft]
\item \emph{Notes.} ***, **, * refers to statistical significance at the 1, 5, and 10\% level. Robust standard errors for estimates of $\beta_Y$ in parentheses. The table shows estimated parameters $\alpha_Y$ and $\beta_Y$ (as presented in equation \ref{eqn:rf}) on secondary school major choice.
\end{tablenotes}
\end{threeparttable}
\end{center}
\end{singlespace}
\end{table}

\newpage
\section{Long term effects}\label{app:long_term_effects}

\begin{table}[hbt!]
\begin{singlespace}
    \begin{center}
    \begin{threeparttable}
    \caption{\label{tab:table_c2016_e4_1} Estimated fractions of students with positive (\emph{Trapped in Track}) and negative (\emph{Slow Starter}) effects of a positive change in first-year secondary school track enrollment, on \textbf{secondary school track enrollment four years after the start of secondary education}. Columns 5–6 report bounds on the difference between these fractions, while Columns 7–8 report bounds on their sum. The results apply to the primary school graduation cohorts of 2014/15 to 2016/17 for which we can report long term effects.}
	\begin{tabularx}{1\textwidth}{@{\extracolsep{\fill}} lcccccccc}
		\toprule
         &\multicolumn{2}{c}{\emph{Trapped in Track}}&\multicolumn{2}{c}{\emph{Slow Starter}}&\multicolumn{2}{c}{Net EFFECT}&\multicolumn{2}{c}{Total EFFECT} \\ 
         \cmidrule(lr){2-5}\cmidrule(lr){6-7}\cmidrule(lr){8-9}
        &LB&UB&LB&UB&LB&UB&LB&UB\\    
            \cmidrule(lr){2-9}
            &(1)&(2)&(3)&(4)&(5)&(6)&(7)&(8)\\
		\cmidrule(lr){2-9}
    \input{Tables2/table_c2016_e4_1_60w.csv}\\
    \input{Tables2/table_c2016_e4_1_52w.csv}\\
    \input{Tables2/table_c2016_e4_1_50w.csv}\\
    \input{Tables2/table_c2016_e4_1_34w.csv}\\
    \input{Tables2/table_c2016_e4_1_30w.csv}\\
    \input{Tables2/table_c2016_e4_1_61h.csv}\\
    \input{Tables2/table_c2016_e4_1_60h.csv}\\
    \input{Tables2/table_c2016_e4_1_52h.csv}\\
    \input{Tables2/table_c2016_e4_1_50h.csv}\\
    \input{Tables2/table_c2016_e4_1_34h.csv}\\
    \bottomrule
	\end{tabularx}
\begin{tablenotes}[flushleft]
\item \emph{Notes.} ***, **, * refers to statistical significance at the 1, 5, and 10\% level. + indices that there is no information on the statistical significance of the estimates. In the next update we plan to compute indicators of statistical significance using the bootstrap method.
\end{tablenotes}
\end{threeparttable}
\end{center}
\end{singlespace}
\end{table}

\begin{table}[!h]
\begin{singlespace}
    \begin{center}
    \begin{threeparttable}
    \caption{\label{tab:table_c2016_gr8_1} Estimated fractions of students with positive (\emph{Trapped in Track}) and negative (\emph{Slow Starter}) effects of a positive change in first-year secondary school track enrollment, on \textbf{highest secondary school track level attained eight years after the start of secondary education}. Columns 5–6 report bounds on the difference between these fractions, while Columns 7–8 report bounds on their sum. The results apply to the primary school graduation cohorts of 2014/15 to 2016/17.}
	\begin{tabularx}{1\textwidth}{@{\extracolsep{\fill}} lcccccccc}
		\toprule
         &\multicolumn{2}{c}{TT}&\multicolumn{2}{c}{SS}&\multicolumn{2}{c}{Net EFFECT}&\multicolumn{2}{c}{Total EFFECT} \\ 
         \cmidrule(lr){2-5}\cmidrule(lr){6-7}\cmidrule(lr){8-9}
        &LB&UB&LB&UB&LB&UB&LB&UB\\    
            \cmidrule(lr){2-9}
            &(1)&(2)&(3)&(4)&(5)&(6)&(7)&(8)\\
		\cmidrule(lr){2-9}
    \input{Tables2/table_c2016_gr8_1_60w.csv}\\
    \input{Tables2/table_c2016_gr8_1_52w.csv}\\
    \input{Tables2/table_c2016_gr8_1_50w.csv}\\
    \input{Tables2/table_c2016_gr8_1_34w.csv}\\
    \input{Tables2/table_c2016_gr8_1_30w.csv}\\
    \input{Tables2/table_c2016_gr8_1_61h.csv}\\
    \input{Tables2/table_c2016_gr8_1_60h.csv}\\
    \input{Tables2/table_c2016_gr8_1_52h.csv}\\
    \input{Tables2/table_c2016_gr8_1_50h.csv}\\
    \input{Tables2/table_c2016_gr8_1_34h.csv}\\
    \bottomrule
	\end{tabularx}
\begin{tablenotes}[flushleft]
\item \emph{Notes.} ***, **, * refers to statistical significance at the 1, 5, and 10\% level. + indices that there is no information on the statistical significance of the estimates. In the next update we plan to compute indicators of statistical significance using the bootstrap method.
\end{tablenotes}
\end{threeparttable}
\end{center}
\end{singlespace}
\end{table}

\begin{table}[!h]
\begin{singlespace}
    \begin{center}
    \begin{threeparttable}
    \caption{\label{tab:table_c2016_gr8_2} Estimated fractions of students with positive (\emph{Trapped in Track}) and negative (\emph{Slow Starter}) effects of a positive change in first-year secondary school track enrollment, on \textbf{highest tertiary school type attained between five and eight years after the start of secondary education}. Columns 5–6 report bounds on the difference between these fractions, while Columns 7–8 report bounds on their sum. The results apply to the primary school graduation cohorts of 2014/15 to 2016/17.}
	\begin{tabularx}{1\textwidth}{@{\extracolsep{\fill}} lcccccccc}
		\toprule
         &\multicolumn{2}{c}{TT}&\multicolumn{2}{c}{SS}&\multicolumn{2}{c}{Net EFFECT}&\multicolumn{2}{c}{Total EFFECT} \\ 
         \cmidrule(lr){2-5}\cmidrule(lr){6-7}\cmidrule(lr){8-9}
        &LB&UB&LB&UB&LB&UB&LB&UB\\    
            \cmidrule(lr){2-9}
            &(1)&(2)&(3)&(4)&(5)&(6)&(7)&(8)\\
		\cmidrule(lr){2-9}
    \input{Tables2/table_c2016_he58_1_60w.csv}\\
    \input{Tables2/table_c2016_he58_1_52w.csv}\\
    \input{Tables2/table_c2016_he58_1_50w.csv}\\
    \input{Tables2/table_c2016_he58_1_34w.csv}\\
    \input{Tables2/table_c2016_he58_1_30w.csv}\\
    \input{Tables2/table_c2016_he58_1_61h.csv}\\
    \input{Tables2/table_c2016_he58_1_60h.csv}\\
    \input{Tables2/table_c2016_he58_1_52h.csv}\\
    \input{Tables2/table_c2016_he58_1_50h.csv}\\
    \input{Tables2/table_c2016_he58_1_34h.csv}\\
    \bottomrule
	\end{tabularx}
\begin{tablenotes}[flushleft]
\item \emph{Notes.} ***, **, * refers to statistical significance at the 1, 5, and 10\% level. + indices that there is no information on the statistical significance of the estimates. In the next update we plan to compute indicators of statistical significance using the bootstrap method.
\end{tablenotes}
\end{threeparttable}
\end{center}
\end{singlespace}
\end{table}

%% file: Tables2/table_arm_e4_1_60w.csv
havo + 2&0.055+&0.078+&0.000+&0.026+&0.038+&0.073+&0.055+&0.098+

%% file: Tables2/table_arm_e4_1_52w.csv
vmbo-gt/havo + 2&0.024+&0.073+&0.000+&0.043+&0.016+&0.042+&0.024+&0.105+

%% file: Tables2/table_arm_e4_1_50w.csv
vmbo-gt + 2&0.026+&0.058+&0.006+&0.049+&-0.001+&0.026+&0.033+&0.105+

%% file: Tables2/table_arm_e4_1_34w.csv
vmbo-kb/gt + 2&0.056+&0.074+&0.007+&0.030+&0.037+&0.054+&0.063+&0.104+

%% file: Tables2/table_arm_e4_1_30w.csv
vmbo-kb + 2&0.069+&0.103+&0.000+&0.059+&0.033+&0.085+&0.069+&0.162+

%% file: Tables2/table_arm_e4_1_61h.csv
havo/vwo + 1&0.041+&0.074+&0.000+&0.017+&0.031+&0.061+&0.041+&0.092+

%% file: Tables2/table_arm_e4_1_60h.csv
havo + 1&0.020+&0.026+&0.000+&0.003+&0.017+&0.022+&0.020+&0.037+

%% file: Tables2/table_arm_e4_1_52h.csv
vmbo-gt/havo + 1&0.052+&0.090+&0.000+&0.039+&0.037+&0.073+&0.052+&0.128+

%% file: Tables2/table_arm_e4_1_50h.csv
vmbo-gt + 1&0.009+&0.017+&0.005+&0.013+&0.004+&0.007+&0.014+&0.042+

%% file: Tables2/table_arm_e4_1_34h.csv
vmbo-kb/gt + 1&0.011+&0.031+&0.006+&0.037+&-0.008+&0.011+&0.017+&0.068+

%% file: Tables2/table_arm_e4_1_30h.csv
vmbo-kb + 1&0.000+&0.041+&0.010+&0.050+&-0.030+&0.011+&0.010+&0.073+

%% file: Tables2/table_rijk_e4_1_60w.csv
havo + 2&0.024+&0.038+&0.001+&0.012+&0.021+&0.028+&0.025+&0.051+

%% file: Tables2/table_rijk_e4_1_52w.csv
vmbo-gt/havo + 2&0.001+&0.005+&0.050+&0.059+&-0.055+&-0.047+&0.051+&0.064+

%% file: Tables2/table_rijk_e4_1_50w.csv
vmbo-gt + 2&0.027+&0.057+&0.001+&0.016+&0.013+&0.041+&0.028+&0.073+

%% file: Tables2/table_rijk_e4_1_34w.csv
vmbo-kb/gt + 2&0.001+&0.077+&0.017+&0.083+&-0.020+&0.003+&0.018+&0.161+

%% file: Tables2/table_rijk_e4_1_30w.csv
vmbo-kb + 2&0.086+&0.096+&0.000+&0.027+&0.067+&0.092+&0.086+&0.123+

%% file: Tables2/table_rijk_e4_1_61h.csv
havo/vwo + 1&0.038+&0.078+&0.000+&0.041+&0.031+&0.052+&0.038+&0.119+

%% file: Tables2/table_rijk_e4_1_60h.csv
havo + 1&0.016+&0.019+&0.000+&0.008+&0.008+&0.017+&0.016+&0.027+

%% file: Tables2/table_rijk_e4_1_52h.csv
vmbo-gt/havo + 1&0.057+&0.073+&0.000+&0.012+&0.045+&0.065+&0.057+&0.085+

%% file: Tables2/table_rijk_e4_1_50h.csv
vmbo-gt + 1&0.001+&0.006+&0.002+&0.007+&-0.003+&0.000+&0.003+&0.019+

%% file: Tables2/table_rijk_e4_1_34h.csv
vmbo-kb/gt + 1&0.011+&0.012+&0.010+&0.010+&0.002+&0.002+&0.021+&0.067+

%% file: Tables2/table_rijk_e4_1_30h.csv
vmbo-kb + 1&0.038+&0.046+&0.009+&0.025+&0.021+&0.029+&0.047+&0.076+

%% file: Tables2/table_rf_60w_baseline_major.csv
havo + 2&0.515&-0.004 (0.012)&0.275&0.004 (0.010)&0.210&0.000 (0.009)

%% file: Tables2/table_rf_52w_baseline_major.csv
vmbo-gt/havo + 2&0.504&-0.014 (0.019)&0.318&0.017 (0.018)&0.178&-0.003 (0.014)

%% file: Tables2/table_rf_50w_baseline_major.csv
vmbo-gt + 2&0.500&-0.024** (0.012)&0.326&0.016 (0.011)&0.174&0.008 (0.009)

%% file: Tables2/table_rf_34w_baseline_major.csv
vmbo-kb/gt + 2&0.565&-0.022 (0.030)&0.243&0.040 (0.026)&0.192&-0.018 (0.023)

%% file: Tables2/table_rf_30w_baseline_major.csv
vmbo-kb + 2&0.600&-0.020 (0.015)&0.202&0.031** (0.012)&0.199&-0.011 (0.012)

%% file: Tables2/table_rf_22w_baseline_major.csv
vmbo-bb/kb + 2&0.633&0.005 (0.060)&0.190&0.002 (0.050)&0.177&-0.007 (0.047)

%% file: Tables2/table_rf_20w_baseline_major.csv
vmbo-bb + 2&0.583&0.005 (0.025)&0.193&-0.002 (0.020)&0.224&-0.003 (0.021)

%% file: Tables2/table_rf_61h_baseline_major.csv
havo/vwo + 1&0.522&-0.008 (0.014)&0.284&-0.013 (0.012)&0.194&0.021* (0.011)

%% file: Tables2/table_rf_60h_baseline_major.csv
havo + 1&0.540&0.000 (0.011)&0.289&0.005 (0.010)&0.170&-0.005 (0.008)

%% file: Tables2/table_rf_52h_baseline_major.csv
vmbo-gt/havo + 1&0.492&0.015 (0.017)&0.343&-0.007 (0.016)&0.165&-0.008 (0.013)

%% file: Tables2/table_rf_50h_baseline_major.csv
vmbo-gt + 1&0.493&0.003 (0.012)&0.340&0.007 (0.011)&0.167&-0.010 (0.009)

%% file: Tables2/table_rf_34h_baseline_major.csv
vmbo-kb/gt + 1&0.595&-0.043 (0.029)&0.261&0.016 (0.026)&0.144&0.027 (0.021)

%% file: Tables2/table_rf_30h_baseline_major.csv
vmbo-kb + 1&0.575&0.034 (0.040)&0.196&-0.003 (0.032)&0.229&-0.031 (0.033)

%% file: Tables2/table_rf_22h_baseline_major.csv
vmbo-bb/kb + 1&0.633&0.001 (0.030)&0.177&0.000 (0.024)&0.189&-0.001 (0.025)

%% file: Tables2/table_rf_20h_baseline_major.csv
vmbo-bb + 1&0.683&-0.050** (0.024)&0.147&0.012 (0.019)&0.170&0.038* (0.020)

%% file: Tables2/table_c2016_e4_1_60w.csv
havo + 2&0.041+&0.059+&0.000+&0.012+&0.029+&0.054+&0.041+&0.066+

%% file: Tables2/table_c2016_e4_1_52w.csv
vmbo-gt/havo + 2&0.006+&0.021+&0.036+&0.053+&-0.042+&-0.021+&0.042+&0.075+

%% file: Tables2/table_c2016_e4_1_50w.csv
vmbo-gt + 2&0.023+&0.049+&0.003+&0.028+&0.001+&0.028+&0.027+&0.076+

%% file: Tables2/table_c2016_e4_1_34w.csv
vmbo-kb/gt + 2&0.068+&0.099+&0.004+&0.034+&0.051+&0.071+&0.071+&0.134+

%% file: Tables2/table_c2016_e4_1_30w.csv
vmbo-kb + 2&0.083+&0.096+&0.000+&0.030+&0.057+&0.093+&0.083+&0.126+

%% file: Tables2/table_c2016_e4_1_61h.csv
havo/vwo + 1&0.020+&0.076+&0.000+&0.044+&0.013+&0.039+&0.020+&0.113+

%% file: Tables2/table_c2016_e4_1_60h.csv
havo + 1&0.008+&0.018+&0.000+&0.009+&0.001+&0.010+&0.008+&0.027+

%% file: Tables2/table_c2016_e4_1_52h.csv
vmbo-gt/havo + 1&0.062+&0.081+&0.000+&0.010+&0.053+&0.071+&0.062+&0.091+

%% file: Tables2/table_c2016_e4_1_50h.csv
vmbo-gt + 1&0.005+&0.011+&0.010+&0.013+&-0.007+&0.000+&0.016+&0.023+

%% file: Tables2/table_c2016_e4_1_34h.csv
vmbo-kb/gt + 1&0.001+&0.055+&0.008+&0.040+&-0.012+&0.019+&0.009+&0.095+

%% file: Tables2/table_c2016_gr8_1_60w.csv
havo + 2&0.027+&0.051+&0.003+&0.026+&0.002+&0.034+&0.030+&0.068+

%% file: Tables2/table_c2016_gr8_1_52w.csv
vmbo-gt/havo + 2&0.027+&0.044+&0.021+&0.049+&-0.016+&0.020+&0.048+&0.089+

%% file: Tables2/table_c2016_gr8_1_50w.csv
vmbo-gt + 2&0.022+&0.043+&0.003+&0.037+&-0.003+&0.024+&0.026+&0.072+

%% file: Tables2/table_c2016_gr8_1_34w.csv
vmbo-kb/gt + 2&0.078+&0.130+&0.000+&0.050+&0.049+&0.115+&0.078+&0.180+

%% file: Tables2/table_c2016_gr8_1_30w.csv
vmbo-kb + 2&0.075+&0.100+&0.002+&0.050+&0.029+&0.084+&0.078+&0.151+

%% file: Tables2/table_c2016_gr8_1_61h.csv
havo/vwo + 1&0.020+&0.081+&0.000+&0.056+&-0.013+&0.055+&0.020+&0.114+

%% file: Tables2/table_c2016_gr8_1_60h.csv
havo + 1&0.018+&0.023+&0.000+&0.009+&0.009+&0.022+&0.018+&0.032+

%% file: Tables2/table_c2016_gr8_1_52h.csv
vmbo-gt/havo + 1&0.034+&0.056+&0.011+&0.032+&0.011+&0.030+&0.045+&0.085+

%% file: Tables2/table_c2016_gr8_1_50h.csv
vmbo-gt + 1&0.005+&0.010+&0.004+&0.014+&-0.006+&0.004+&0.009+&0.024+

%% file: Tables2/table_c2016_gr8_1_34h.csv
vmbo-kb/gt + 1&0.000+&0.043+&0.020+&0.058+&-0.039+&0.002+&0.020+&0.102+

%% file: Tables2/table_c2016_he58_1_60w.csv
havo + 2&0.019+&0.044+&0.000+&0.013+&0.012+&0.032+&0.019+&0.057+

%% file: Tables2/table_c2016_he58_1_52w.csv
vmbo-gt/havo + 2&0.005+&0.030+&0.024+&0.050+&-0.022+&-0.019+&0.029+&0.080+

%% file: Tables2/table_c2016_he58_1_50w.csv
vmbo-gt + 2&0.015+&0.043+&0.000+&0.032+&0.001+&0.028+&0.015+&0.072+

%% file: Tables2/table_c2016_he58_1_34w.csv
vmbo-kb/gt + 2&0.033+&0.124+&0.000+&0.102+&-0.023+&0.067+&0.033+&0.191+

%% file: Tables2/table_c2016_he58_1_30w.csv
vmbo-kb + 2&0.014+&0.098+&0.000+&0.080+&-0.025+&0.040+&0.014+&0.160+

%% file: Tables2/table_c2016_he58_1_61h.csv
havo/vwo + 1&0.012+&0.046+&0.020+&0.067+&-0.036+&0.012+&0.033+&0.108+

%% file: Tables2/table_c2016_he58_1_60h.csv
havo + 1&0.016+&0.026+&0.000+&0.005+&0.016+&0.024+&0.016+&0.030+

%% file: Tables2/table_c2016_he58_1_52h.csv
vmbo-gt/havo + 1&0.017+&0.036+&0.000+&0.031+&-0.002+&0.018+&0.017+&0.067+

%% file: Tables2/table_c2016_he58_1_50h.csv
vmbo-gt + 1&0.023+&0.024+&0.000+&0.000+&0.023+&0.024+&0.023+&0.024+

%% file: Tables2/table_c2016_he58_1_34h.csv
vmbo-kb/gt + 1&0.000+&0.046+&0.029+&0.082+&-0.056+&-0.008+&0.029+&0.112+